\documentclass[aps,prd,12pt,notitlepage,nofootinbib,tightenlines]{revtex4-1}
\usepackage{amsmath}
\usepackage{bm}
\usepackage{times}
\usepackage{braket}
\usepackage{color}
\usepackage{epsfig}
\usepackage{slashed}
\usepackage{hyperref}
\usepackage{subfigure}
\usepackage{graphics}
\allowdisplaybreaks[1]
\definecolor{nicered}{rgb}{0.7,0.1,0.1}
\definecolor{nicegreen}{rgb}{0.1,0.5,0.1}

\definecolor{diff}{rgb}{0,0,0}

\hypersetup{colorlinks,citecolor=nicegreen,linkcolor=nicered}

\begin{document}
	
	\title{Branching fractions of $B^-\to D^-X_{0,1}(2900)$ and their implications}
	
	\author{Yan-Ke Chen$^{1}$}\email{chenyk16@lzu.edu.cn}
	\author{Jia-Jie Han$^{2}$}\email{hanjiajie1020@163.com}
	\author{Qi-Fang L\"u$^{3}$}\email{lvqifang@hunnu.edu.cn}
	\author{Jian-Peng Wang$^{1}$}\email{jpwang2016@lzu.edu.cn}
	\author{Fu-Sheng Yu$^{1}$}\email{yufsh@lzu.edu.cn}
	
	\affiliation{$^1$ School of Nuclear Science and Technology, Lanzhou University, Lanzhou 730000, China\\
	$^2$ Department of Physics and Institute of Theoretical Physics, Nanjing Normal University, Nanjing, Jiangsu 210023, China\\
	$^3$ Department of Physics, Hunan Normal University, Changsha 410081, China}
	
	
	\begin{abstract}
	The exotic states $X_{0,1}(2900)$ with the quark flavor of $cs\bar{u}\bar{d}$ are recently observed in the mass spectrum of $D^+K^-$ in $B^-\to D^-D^+K^-$ by the LHCb collaboration. To explore the nature of $X_{0,1}(2900)$, except for analyzing their masses and decay widths as usually did in literatures, the study of their production mechanism in $B$-meson weak decays would provide another important information. The amplitude of $B^-\to D^- X_{0,1}$ is non-factorizable. We consider the final-state-interaction effects and calculate them via the rescattering mechanism. The measured branching fractions of $B^-\to D^- X_{0,1}$ are revealed. It is manifested by ${B}^-\to \Lambda_c^-\Xi_c^{(\prime)0}$ and $\Lambda_b^0\to P_c^+K^-$ that the rescattering mechanism can result in the relatively large branching fractions. The similar processes of $B^-\to \pi^-X_{0,1}$ are also analyzed. The isospins of $X_{0,1}$ can be investigated by  $B\to DX_{0,1}^{\pm,0}$ decays.
	\end{abstract}
	
	
	
	\vspace{1cm}
	\maketitle
	
	\section{Introduction}\label{sec:intro}
	
	Very recently, the LHCb collaboration reported the discovery of two new exotic structures $X_0(2900)$ and $X_1(2900)$ in the $D^+ K^-$ invariant mass distributions of the decay process of $B^- \to D^+ D^- K^-$ \cite{LHCb1,LHCb2}.
 Their spin-parity quantum numbers are $J^P=0^+$ and $1^-$, respectively, with the masses and widths as $M_{X_0(2900)}=2866\pm 7{\rm MeV},~\Gamma_{X_0(2900)}=57\pm13{\rm MeV},~M_{X_1(2900)}=2904\pm 5{\rm MeV},~\Gamma_{X_1(2900)}=110\pm12{\rm MeV}$. A lot of theoretical works are proposed to explain these two exotic states as either compact tetraquarks or molecular states, or non-resonant triangle singularities \cite{Karliner:2020vsi,Hu:2020mxp,He:2020jna,Liu:2020orv,Zhang:2020oze,Lu:2020qmp,Liu:2020nil,Chen:2020aos,He:2020btl,Wang:2020xyc,Huang:2020ptc,Qin:2020zlg,Xue:2020vtq,Molina:2020hde,Burns:2020epm,Agaev:2020nrc,Albuquerque:2020ugi}.
The $D^+ K^-$ channel implies those two states are composed by four fully different flavors, $cs\bar u\bar d$. The open-heavy-flavor states with four different flavors are extremely exotic and helpful for understanding the low-energy non-perturbation behavior of the QCD and nature of the strong interactions, thus of high interests both in the theoretical studies and experimental searches \cite{Yu:2017pmn,Xing:2019hjg,Chen:2018hts,Huang:2019otd,Cheng:2020nho,Liu:2016ogz,He:2016xvd,D0:2016mwd,Aaij:2016iev,Molina:2010tx}.
	
	To explore the nature of exotic states, the production mechanism would provide  important information, except for analyzing the masses and decay widths as usually did in literatures. It was proposed to search for the $bs\bar u\bar d$ and $cs\bar u\bar d$ tetraquarks in the direct production in the hadron collisions \cite{Yu:2017pmn}. Constituting with only one heavy quark, these tetraquarks are expected to be produced with a large rate. The lowest thresholds are $BK$ or $DK$ which are 270 MeV higher than $B_s\pi$ or $D_s\pi$ of other four-different-flavor tetraquark states. There are some possibilities for the mass regions of the lowest-lying ground states of $bs\bar u\bar d$ or $cs\bar u\bar d$ below the $BK$ or $DK$ thresholds \cite{Yu:2017pmn,Xing:2019hjg,Chen:2018hts,Huang:2019otd,Cheng:2020nho,Liu:2016ogz,He:2016xvd}. In these cases, these states are stable and can only decay weakly. The longer lifetimes will be helpful for the experimental searches in the hadron colliders by rejecting the backgrounds from the primary vertex \cite{Yu:2017pmn}. For the observed $X_{0,1}(2900)$, their masses are much higher than the $DK$ threshold, indicating that $X_{0,1}$ are not the lowest-lying ground states. They are produced in the $B$-meson weak decays, $B^-\to D^- X_{0,1}$  \cite{LHCb1,LHCb2}. From Table. \ref{table:1}, the relatively large branching fractions are the key point in the observation of $X_{0,1}(2900)$. Therefore, it is necessary to understand the production mechanism of $X_{0,1}(2900)$ in the weak decays of $B$ mesons and the corresponding branching fractions. 
		\begin{table}[tbp]
		\caption{The fit fractions of $B^-\to D^-X_{0,1}(2900)$ in $B^- \to D^+ D^- K^-$ and the corresponding branching fractions obtained by LHCb \cite{LHCb1,LHCb2}.}
		\begin{tabular}{ccc}
			\hline
			\hline
			&\ \ \ \ \ \ \ \ Fit fraction\ \ \ \ \ \ \ \  &$Br_{\rm exp}$\\
			\hline
			$B^-\to D^-X_0(2900)^0$&$(5.6\pm 0.5)\%$&$(1.23\pm 0.41)\times10^{-5}$\\
			\hline
			$B^-\to D^-X_1(2900)^0$&$(30.6\pm 3.2)\%$&$(6.73\pm 2.26)\times10^{-5}$\\
			\hline
			\hline
		\end{tabular}
	\label{table:1}
	\end{table}

	\begin{figure}[htbp]
		\centering
		\includegraphics[scale=0.25]{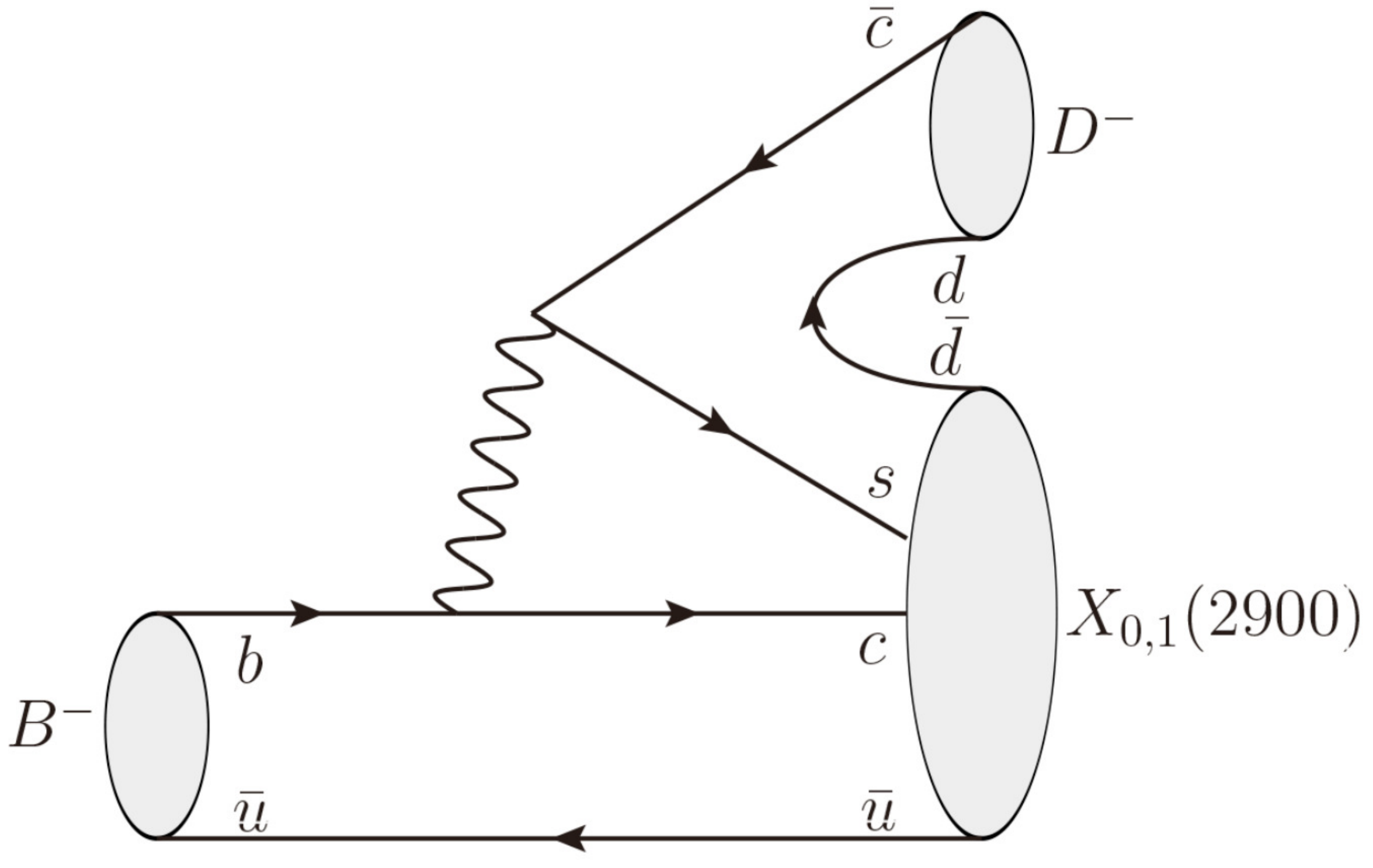}
		\caption{The topological diagram of the $B^-\to D^-X_{0,1}(2900)$ decays.}
		\label{fig:1DXtopo}
	\end{figure}
	The topological diagram of the weak decays of $B^-\to D^-X_{0,1}(2900)$ is shown in Fig. \ref{fig:1DXtopo}. It is an external $W$-emission diagram. Unlike the $B$-meson weak decays into the ordinary quark-antiquark mesons, this diagram is non-factorizable since the weak-interaction produced $s\bar c$ enter into different final states, with additional $d\bar d$ generated from vacuum.  It is not easy to calculate this diagram with the unknown structure of $X_{0,1}(2900)$ in the QCD-inspired methods. With one more quark-antiquark pair in the final state, at least two gluons are required making the gluons not as hard as the ordinary $B$ decays.Therefore, the long-distance contributions would play a significant role in the decays of $B^-\to D^-X_{0,1}$.We would calculate the final-state-interaction (FSI) effects for the production of $X_{0,1}(2900)$ in $B^-\to D^-X_{0,1}$.
		
	The rescattering mechanism for the FSI effects has been successfully applied to predict the discovery channel of the doubly charmed baryon, $\Xi_{cc}^{++}\to \Lambda_c^+K^-\pi^+\pi^+$ \cite{Yu:2017zst,Aaij:2017ueg}, where the long-distance contributions are large in the charm scale. At the bottom scale, it has been tested in the $B\to \pi\pi$, $K\pi$ and $D\pi$ modes \cite{Cheng:2004ru}.Since one more quark-antiquark pair contributes to the decays of $B^-\to D^-X_{0,1}$, we will test the rescattering mechanism again in the processes of $B^-\to \Lambda_c^- \Xi_c^{(\prime)0}$ and $\Lambda_b^0\to P_c^+K^-$ which are both non-factorizable for the similar external or internal $W$-emission diagrams with additional quark-antiquark pair generated from vacuum.  
	
	The existence and the nature of $X_{0,1}(2900)$ should be cross-checked by other processes. The decay of $B^-\to \pi^-X_{0,1}$ is very similar to $B^-\to D^-X_{0,1}$ but with a replacement of $\bar c$ in $D^-$ by $\bar u$ in $\pi^-$. Prediction on the branching fractions of $B^-\to \pi^-X_{0,1}$ would be helpful to confirm the existence of $X_{0,1}(2900)$. Besides, the isospins of  $X_{0,1}(2900)$ are unknown from the current measurement. We propose to measure more isospin-related processes, such as $B^-\to \overline D^0 X_{0,1}^-$, $\overline B^0\to \overline D^0 X_{0,1}^0$ and $\overline B^0\to D^- X_{0,1}^+$. Once the isospin partners $X_i^\pm$ were observed, $X_{i}$ could be determined as isospin triplet.  
	
	This manuscript is organized as follows. In Section \ref{sec:BtoDX}, we calculate the  branching fractions of $B^- \to D^-X_{0,1}$ using the rescattering mechanism, revealing the experimental measurement. The rescattering mechanism for such unusual processes is manifested in Section \ref{sec:BtoLambdaXi} by $B^-\to \Lambda_c^-\Xi_c^{(\prime)0}$ and $\Lambda_b^0\to P_c^+K^-$. The branching fraction of $B^-\to\pi^-X_{0,1}$ are predicted in Section \ref{sec:BtopiX}. The isospin analysis on $B\to DX_{0,1}$ are provided in Section \ref{sec:Isospin}. Section \ref{sec:Summary} is a summary.
	
	\section{Branching fractions of $B^-\to D^-X_{0,1}$}\label{sec:BtoDX}
	The exotic states of $X_{0,1}(2900)$ are observed in the amplitude analysis of $B^-\to D^-D^+K^-$ \cite{LHCb1,LHCb2}. The fit fractions of $B^-\to D^-X_{0,1}$ are as large as 5.6\% and 30.6\%, respectively, shown in Table \ref{table:1}. Considering $Br(B^-\to D^-D^+K^-)=(2.2\pm0.7)\times 10^{-4}$ \cite{PDG}, the branching fractions of $B^-\to D^-X_{0,1}(2900)$ can be obtained and listed in Table \ref{table:1}. The relatively large branching fractions at the order of $10^{-5}$ are important for the experimental measurement and observation. The deep understanding of the production mechanism of $X_{0,1}(2900)$ in the decay of $B^-\to D^-X_{0,1}$ will be helpful for the exploration the nature of $X_{0,1}$ states.

	The topological diagram of $B^-\to D^-X_{0,1}$ decay modes is depicted in Fig \ref{fig:1DXtopo}. This topological diagram induced by $W$-external emission diagram is unusual compared to ordinary $B$ meson decays, since $q\bar{q}^\prime$ pair enter into different final states instead of annihilation. Such a topological diagram are totally non-factorizable and unavailable in the perturbative QCD calculations. On the contrary, these processes are dominated by the long-distance contributions. In this paper, we calculate the long-distance contributions by FSIs effects, which is modeled as soft rescattering of two intermediate particles. The FSIs are usually calculated at hadron level under the one-particle-exchange model, which has achieved great success, especially in prediction of the weak decays of doubly-charmed baryon $\Xi_{cc}^{++}$ \cite{Yu:2017zst}. The applications  of the rescattering mechanism in the weak decays of $B$ and $D$ mesons are given in Refs.\cite{Cheng:2004ru,Lu:2005mx,Li:2002pj}.
	
\begin{figure}[htbp]
		\centering
		\subfigure[]{
			\begin{minipage}[]{0.45\linewidth}
				\includegraphics[scale=0.2]{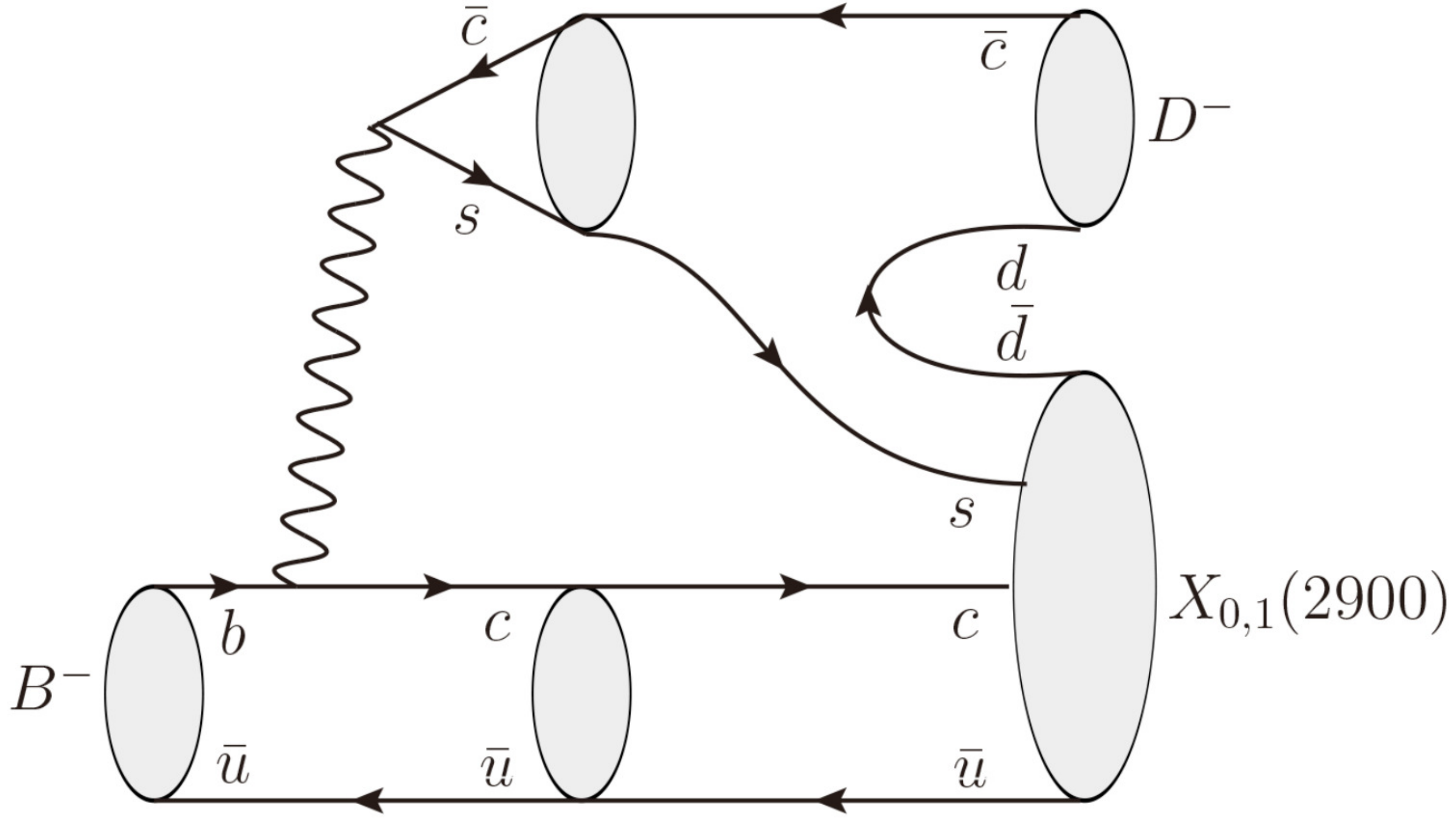}
				\label{fig:2a}
			\end{minipage}
		}
		\subfigure[]{
			\begin{minipage}[]{0.45\linewidth}
				\includegraphics[scale=0.2]{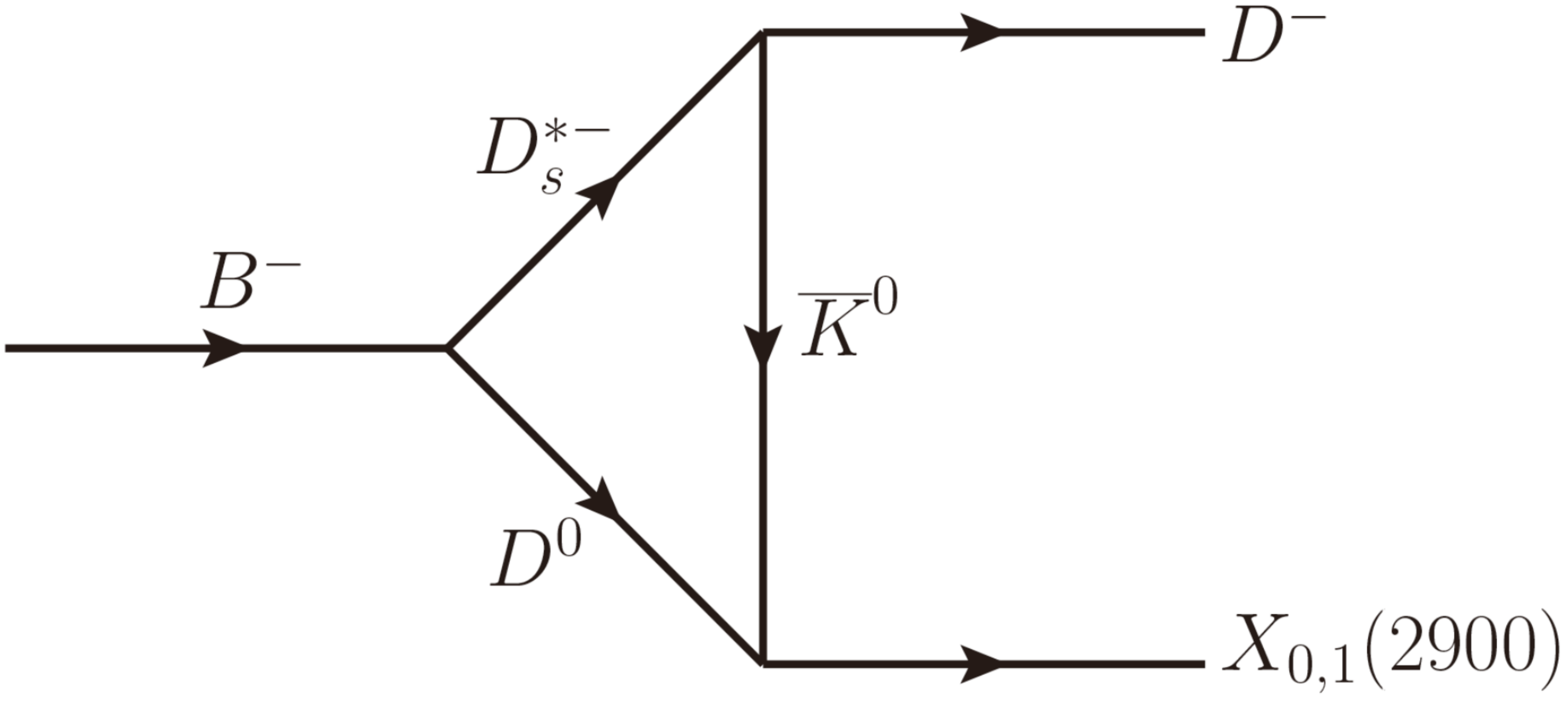}
				\label{fig:2b}
			\end{minipage}
		}
	\caption{(a) Quark-level diagram of $B^-\to D^-X_{0,1}(2900)$ within the rescattering mechanism. (b) The corresponding hadron-level triangle diagram.}
	\label{fig:2}
	\end{figure}

	In the framework of rescattering mechanism, the decay $B^-\to D^-X_{0,1}(2900)$ can most likely proceed as $B^-\to D_s^{\ast -}D^0\to D^-X_{0,1}(2900)$ via exchanging one intermediate state of $\overline{K}^0$. The quark-level diagram is depicted in Fig.\ref{fig:2a}, while the corresponding hadron-level diagram is in Fig.\ref{fig:2b}.
	
	The calculation of the FSIs effect can be carried on in different ways. In general ,the absorptive part of a two-body decay $B^-\to D^-X_{0,1}(2900)$ can be related to a weak decay process $B^-\to D_s^{\ast -}D^0$, followed by the strong rescattering of $D_s^{\ast -}D^0\to D^-X_{0,1}(2900)$. In this work, we assume the absorptive part is dominating and neglect the dispersive part.
	
	Under the factorization approach, the weak-decay vertex can be expressed as:
	\begin{equation}
		\langle D_s^{\ast -}D^0 |\mathcal{H}_{eff} |B^-\rangle=\frac{G_F}{\sqrt{2}}V_{cb}V_{cs}^\ast a_1 f_{D_s^{\ast}}m_{D_s^{\ast}}F_1^{B\to D}(m_{D_s^{\ast}}^2)\;(2\epsilon^\ast_{D_s^{\ast -}}\cdot p_{B^-}),
	\end{equation}
	where $G_F=1.166\times10^{-5}$GeV$^{-2}$ is the Fermi coupling constant, $V_{cb}$ and $V_{cs}$ are the CKM matrix elements, $a_1$ is the effective Wilson coefficient, and $f_{D_s^{\ast}}$ is decay constant of $D_s^{\ast}$ meson. The form factor $F_1^{B\to D}(m_{D_s^{\ast}}^2)$ can be parameterized as $F(q^2)={F(0)}/(1-a \frac{q^2}{m_B^2}+b \frac{q^4}{m_B^4})$, where the parameters $F(0),\ a,\ b$ can be found in Ref.\cite{Cheng:2004ru}.
	
	The absorptive part of the decay amplitude for Fig.\ref{fig:2b} can be written as:
	\begin{equation}
		\begin{aligned}
		\mathcal{A}bs(B^-\to D^-X_0(2900))=&-2i\frac{G_F}{\sqrt{2}}V_{CKM}a_1\int\frac{|\vec{p}_{D_s^{\ast-}}|d\cos\theta d\phi}{32\pi^2m_B}g_{D_s^\ast DK}g_{DKX_0}m_{X_0}\frac{F^2(t,m_K)}{t-m_K^2}\\
		&\cdot f_{D_s^{\ast -}}m_{D_s^{\ast -}}F_1^{B\to D}(M_{D_s^{\ast -}}^2)(p_{D^0}\cdot p_{D^-}-\frac{(p_{D_s^{\ast -}}\cdot p_{D^-})(p_{D_s^{\ast -}}\cdot p_{D^0})}{m_{D_s^{\ast -}}^2}),
		\end{aligned}
	\end{equation}
	\begin{equation}
		\begin{aligned}
		\mathcal{A}bs(B^-\to D^-X_1(2900))=&2i\frac{G_F}{\sqrt{2}}V_{CKM}a_1\int\frac{|\vec{p}_{D_s^{\ast -}}|d\cos\theta d\phi}{32\pi^2m_B}g_{D_s^\ast DK}g_{DKX_1}m_{X_0}\frac{F^2(t,m_K)}{t-m_K^2}\\
		&\cdot f_{D_s^{\ast -}}m_{D_s^{\ast -}}F_1^{B\to D}(M_{D_s^{\ast -}}^2)(p_{D^0}\cdot p_{D^-}-\frac{(p_{D_s^{\ast -}}\cdot p_{D^-})(p_{D_s^{\ast -}}\cdot p_{D^0})}{m_{D_s^{\ast -}}^2})\\
		&\cdot (p_K\cdot \epsilon_{X_1}),
		\end{aligned}
	\end{equation}
	where $t=p_K^2$ and $F(t,m)=(\Lambda^2-m_K^2)/(\Lambda^2-t)$ with 
	\begin{equation}
	\Lambda=m_K+\eta\Lambda_{\rm QCD},
	\end{equation}
$\Lambda_{\rm QCD}=218$MeV. 
			\begin{figure}[bp]
		\centering
		\includegraphics[scale=0.25]{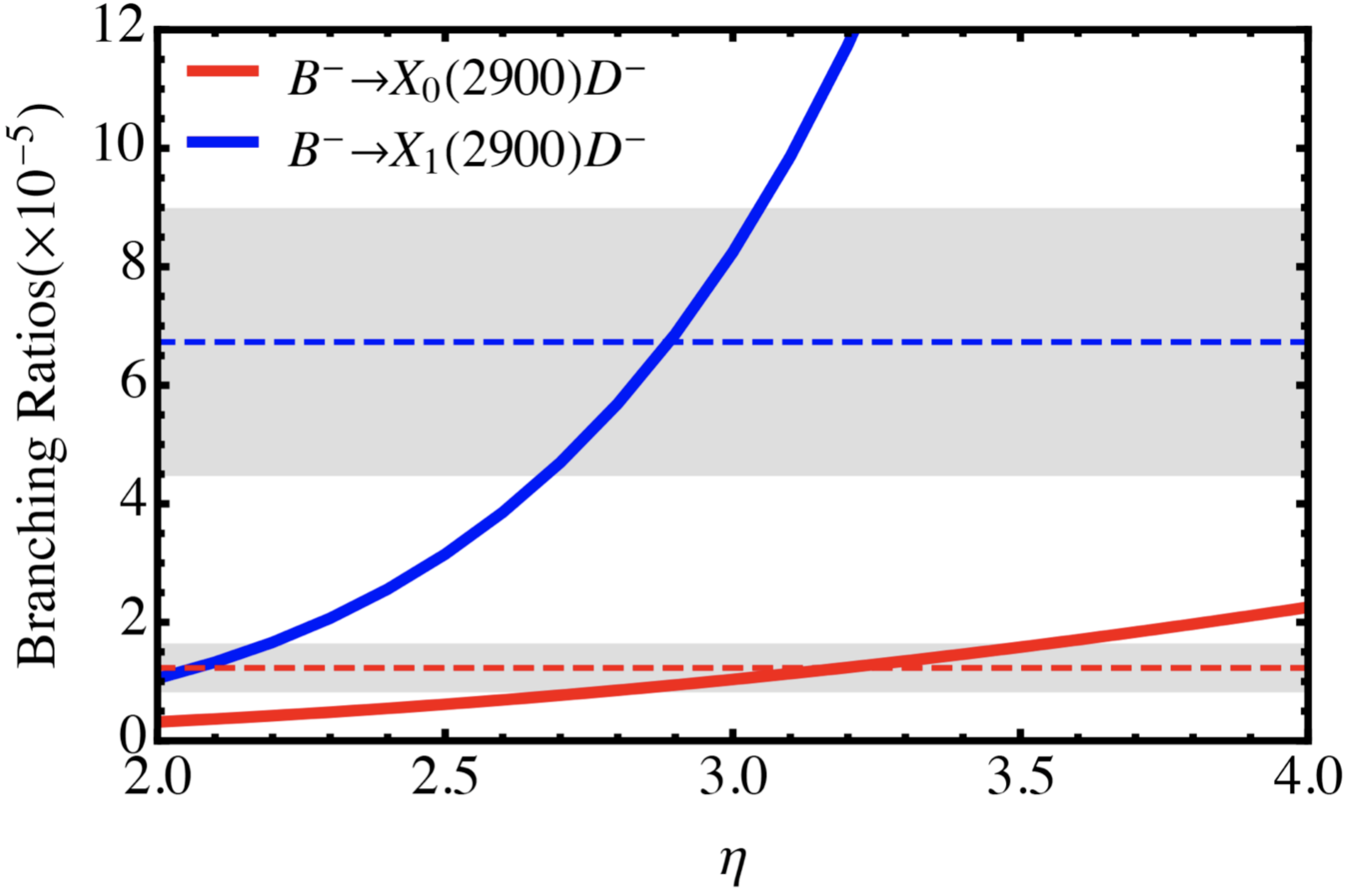}
		\caption{The predictions on the branching fractions of $B^-\to D^-X_{0,1}$ are given by the solid curves with $\eta$ varying from 2.0 to 4.0. The experimental results are shown in dashed lines with the errors indicated by the shadow.}
		\label{fig:3}
	\end{figure}
It will be seen that the results are very sensitive to the values of $\eta$. In principle, there are some other triangle diagrams with the intermediate $D^0\overline K^0$ replaced by $D^{*0}\overline K^{*0}$, $D_1^0\overline K^0$ for $X_0$ and $D^0\overline{K}^{\ast 0}$, $D^{\ast 0}\overline{K}^0$, $D^{\ast 0}\overline{K}^{\ast 0}$, $D_1^0\overline{K}^0$ for $X_1$. Due to the unknown structure of $X_{0,1}(2900)$, the required couplings of $g_{X_{0,1}D^\ast K^\ast}$ or $g_{X_{0,1}D_1K}$ are unknown. Therefore, we neglect such diagrams which doesn't affect our result, since our purpose is to see whether the large branching fractions of $B^-\to D^-X_{0,1}(2900)$ can be understood by the rescattering mechanism. 
	
	The numerical results of the branching fractions of $B^-\to D^-X_{0,1}(2900)$ are provided in Fig.\ref{fig:3}. The red-dashed line and the blue-dashed line are experimental branching fractions for $B^-\to D^-X_0$ and $B^-\to D^-X_1$, respectively. Light-grey shadow areas are experimental error for each decay mode. It can be seen that the theoretical predictions on the branching fractions at around $\eta\approx 3.0$ are consistent with the experimental measurements. The production of $X_{0,1}(2900)$ in $B^-\to D^-X_{0,1}(2900)$ can be understood by the rescattering mechanism. 

	
	\section{Rescattering mechanism in $B^-\to \overline{\Lambda}_c^-\Xi_c^{(\prime)0}$ and $\Lambda_b^0\to P_c^+K^-$}\label{sec:BtoLambdaXi}
\begin{figure}[htbp]
		\centering
		\subfigure[]{
			\begin{minipage}[]{0.3\linewidth}
				\includegraphics[scale=0.22]{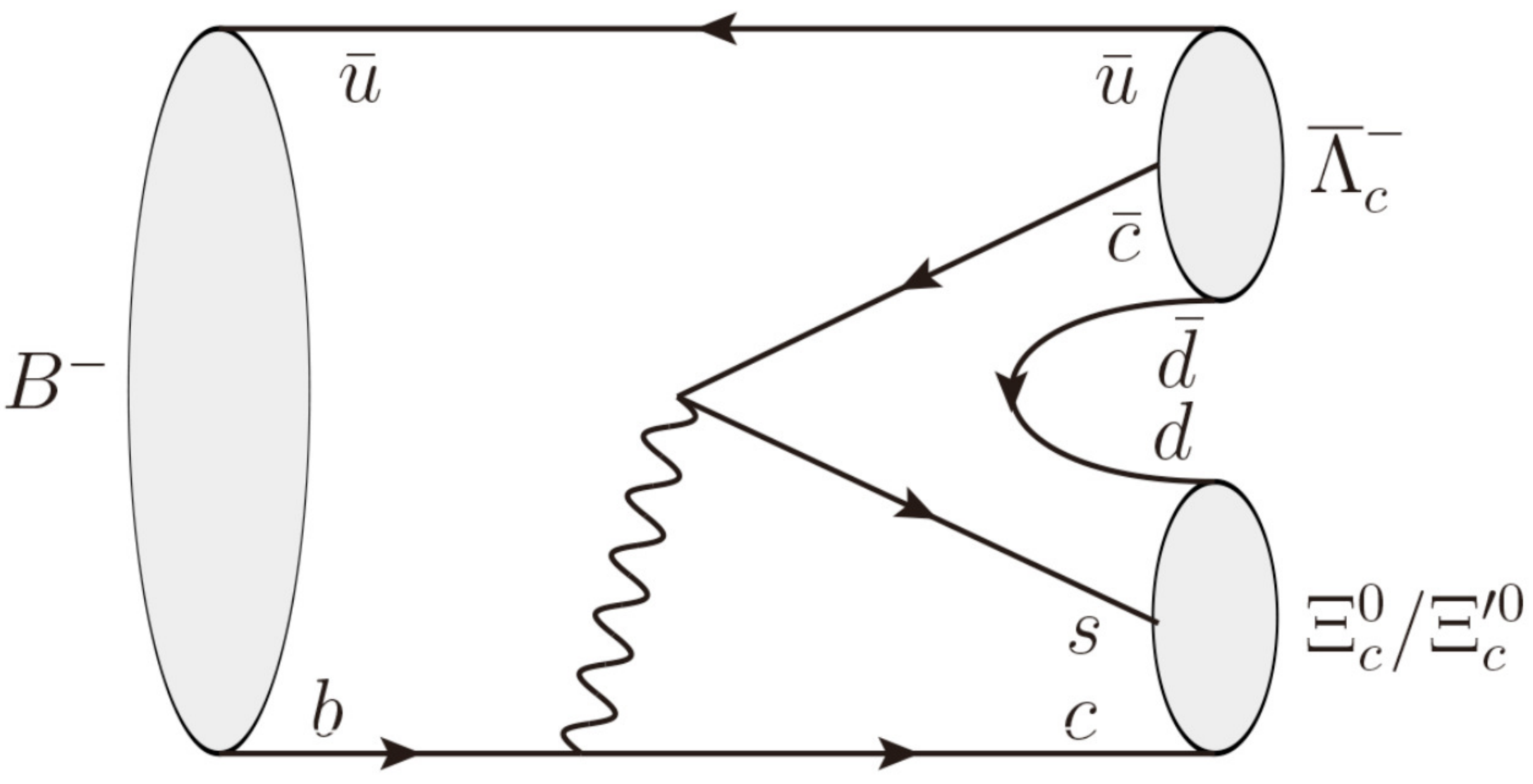}
				\label{fig:4a}
			\end{minipage}
		}
		\subfigure[]{
			\begin{minipage}[]{0.33\linewidth}
				\includegraphics[scale=0.22]{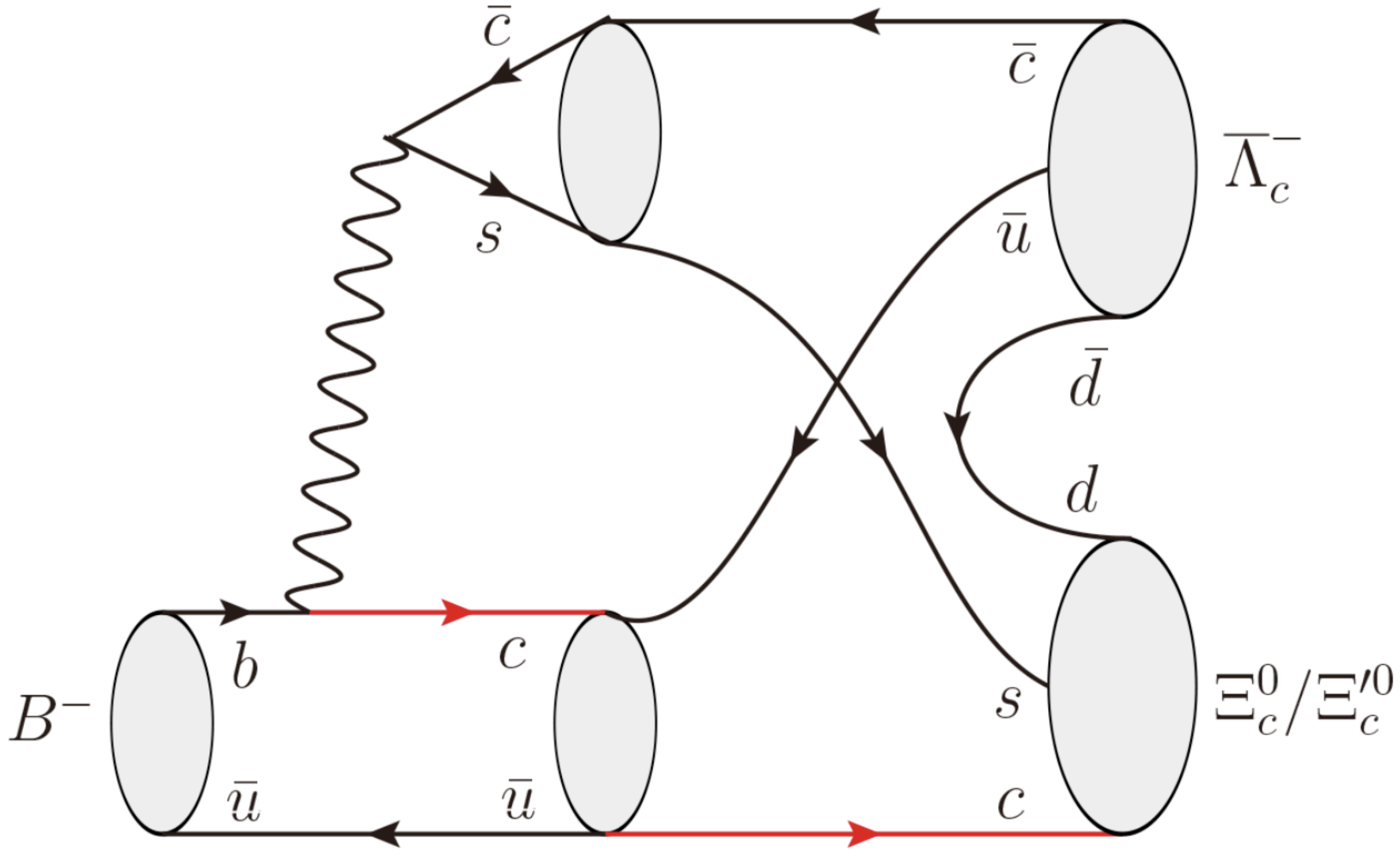}
				\label{fig:4b}
			\end{minipage}
		}
		\subfigure[]{
			\begin{minipage}[]{0.3\linewidth}
				\includegraphics[scale=0.19]{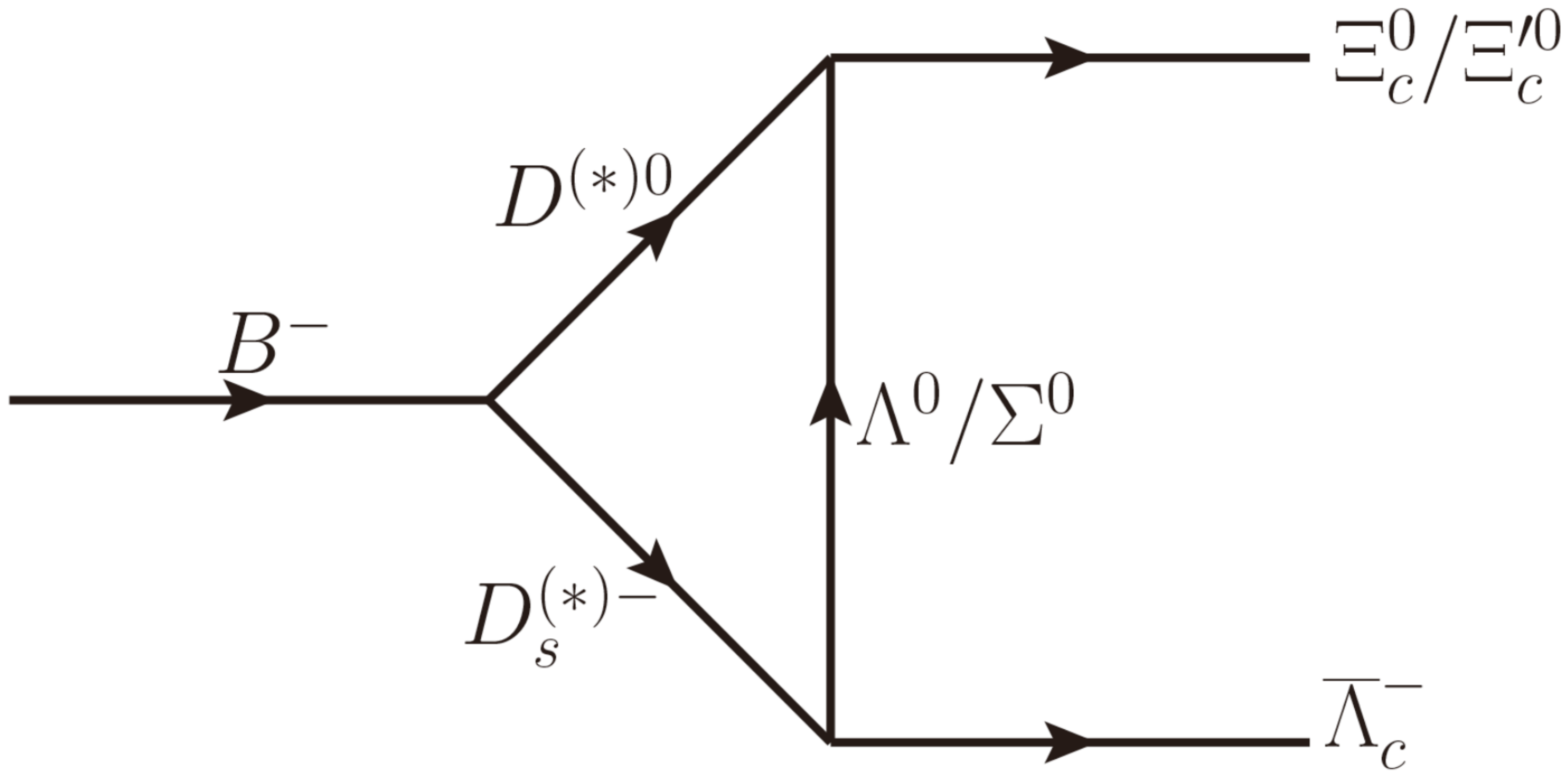}
				\label{fig:4c}
			\end{minipage}
		}
		\caption{The topological diagram, the quark-level diagram and the hadron-level triangle diagram of $B^-\to \overline{\Lambda}_c^-\Xi_c^{(\prime)0}$ within rescattering mechanism are given from the left to the right, respectively. }
		\label{fig:4}
	\end{figure}

\begin{figure}[htbp]
		\centering
		\subfigure[]{
			\begin{minipage}[]{0.3\linewidth}
				\includegraphics[scale=0.4]{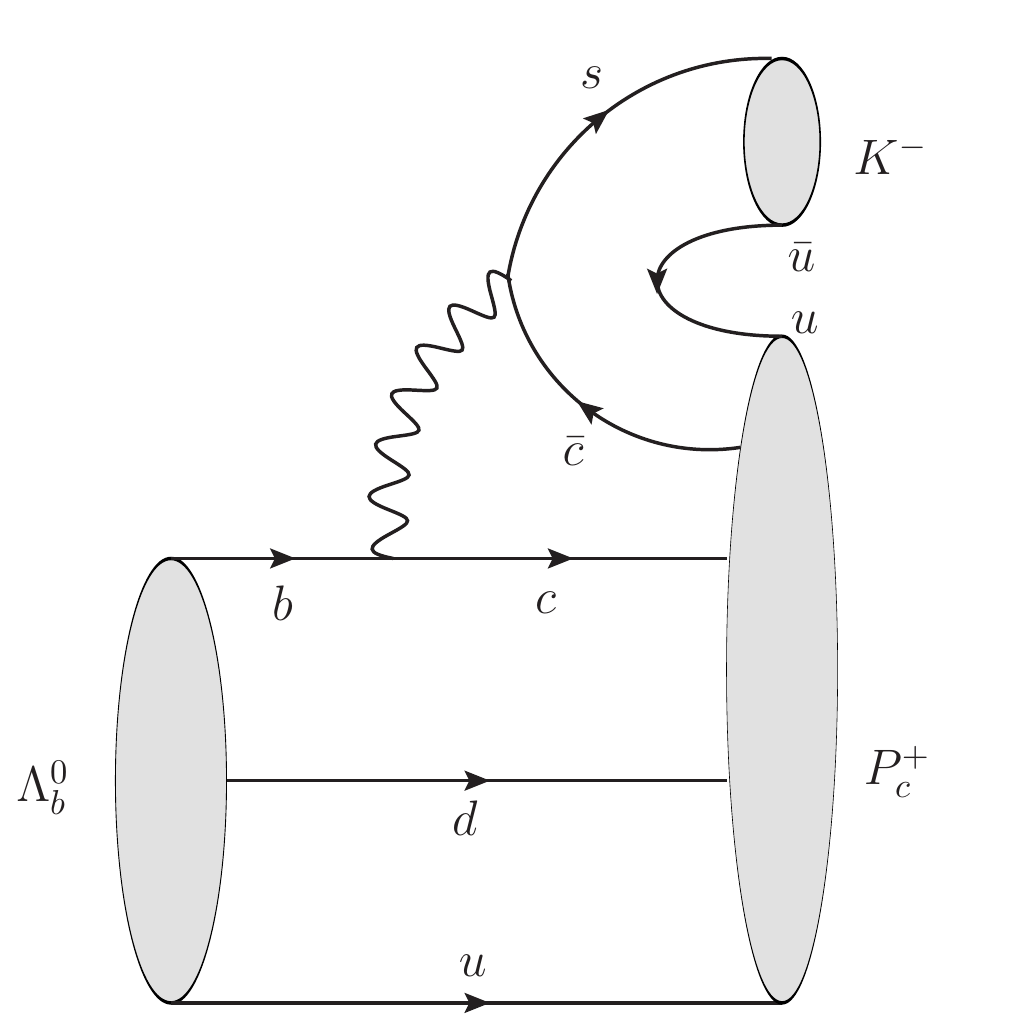}
				\label{fig:5a}
			\end{minipage}
		}
		\subfigure[]{
			\begin{minipage}[]{0.35\linewidth}
				\includegraphics[scale=0.4]{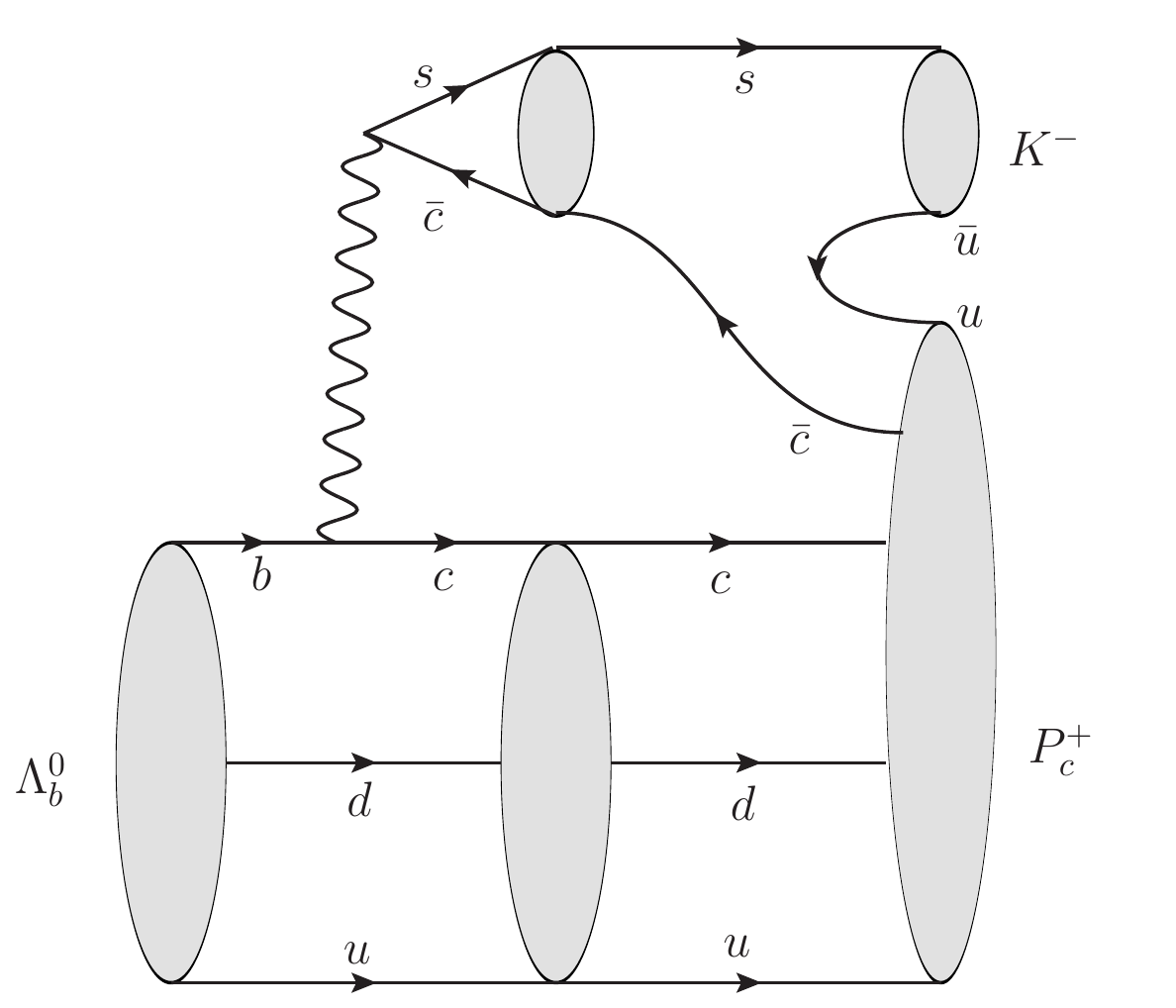}
				\label{fig:5b}
			\end{minipage}
		}
		\subfigure[]{
			\begin{minipage}[]{0.25\linewidth}
				\includegraphics[scale=0.47]{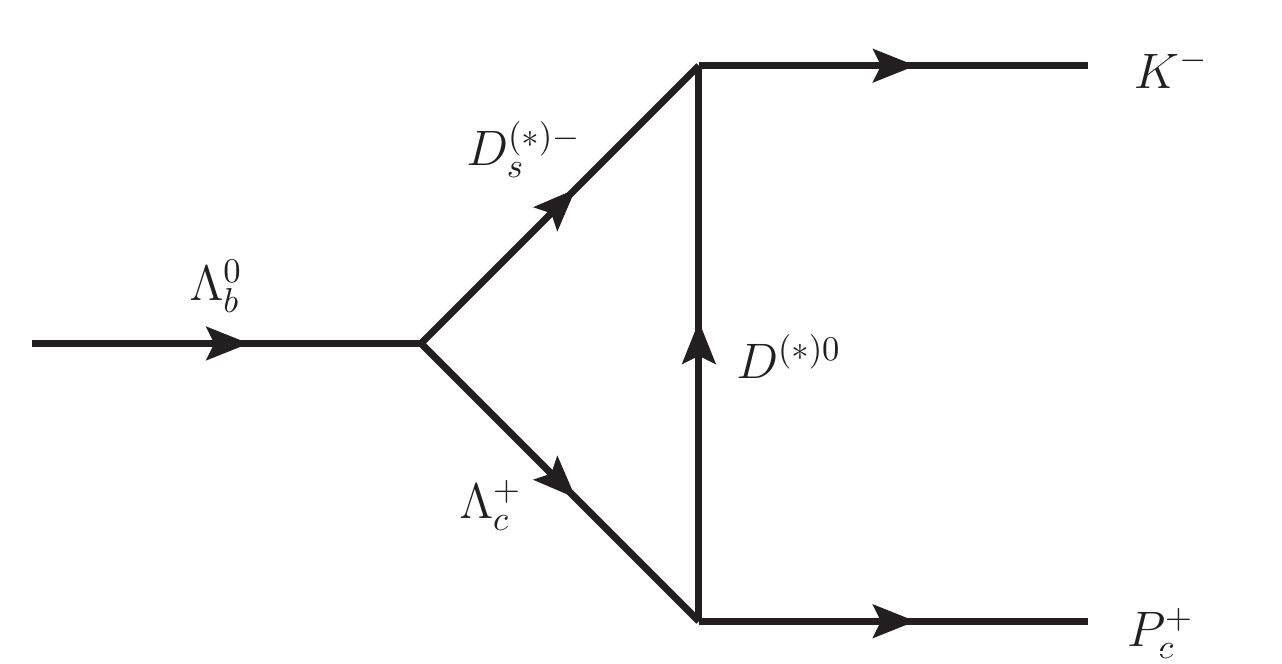}
				\label{fig:5c}
			\end{minipage}
		}
		\caption{Similar to Fig.\ref{fig:4} but for $\Lambda_b^0\to P_c^+K^-$. }
		\label{fig:5}
	\end{figure}

	In the above section, the experimental results can be revealed by the rescattering mechanism. It deserves to check more processes with complicated quark diagrams on the rescattering mechanism. The baryonic decays of $B^-\to \overline{\Lambda}_c^-\Xi_c^{(\prime)0}$ and the pentaquark productions in $\Lambda_b^0\to P_c^+K^-$ have similar topological diagrams with those of $B^-\to D^-X_{0,1}(2900)$. All of them are the external or internal $W$-emission diagrams with a pair of quark and antiquark generated from the vacuum and entering into the different final states. They are non-factorizable and thus calculated by the rescattering mechanism. The topological, quark-level and the hadron-level diagrams of $B^-\to \overline{\Lambda}_c^-\Xi_c^{(\prime)0}$ and $\Lambda_b^0\to P_c^+K^-$ are depicted in Fig.\ref{fig:4} and \ref{fig:5}, respectively. 
	
\begin{figure}[htbp]
		\centering
		\includegraphics[scale=0.57]{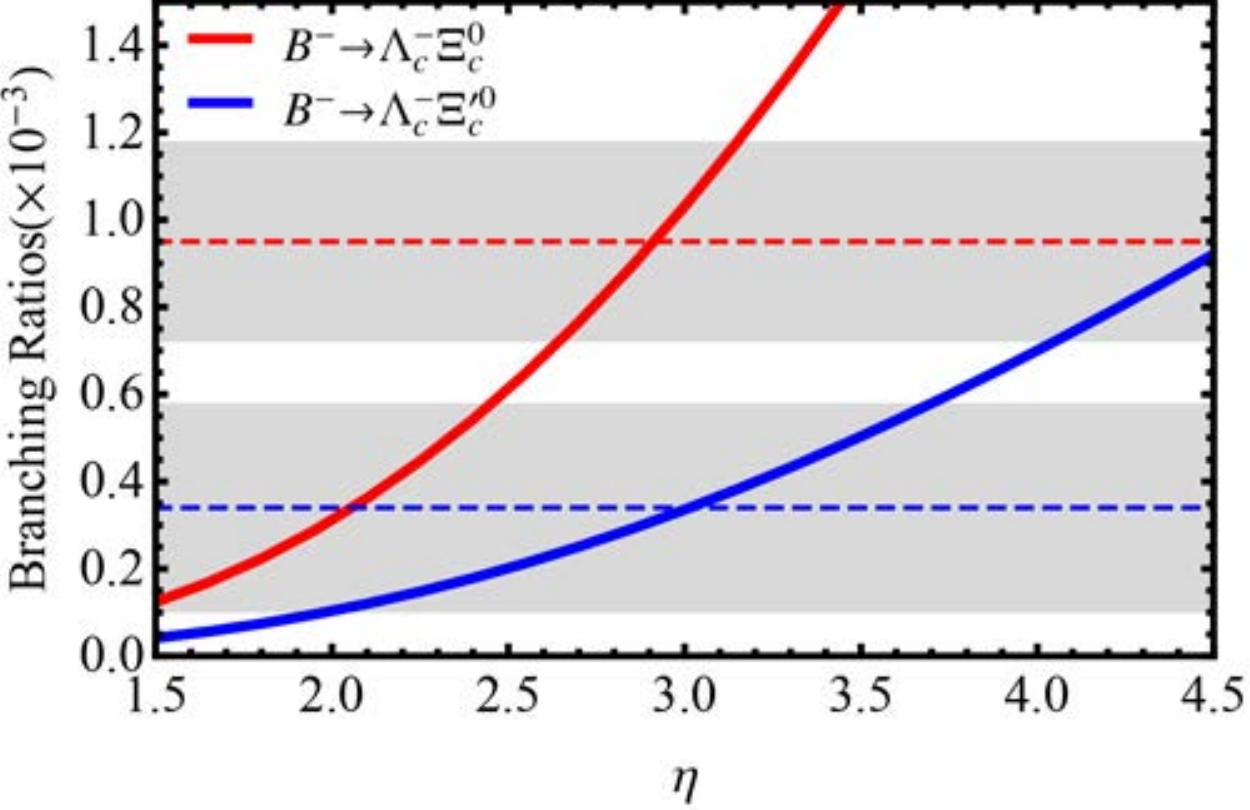}\hspace{1cm}
		\includegraphics[scale=0.37]{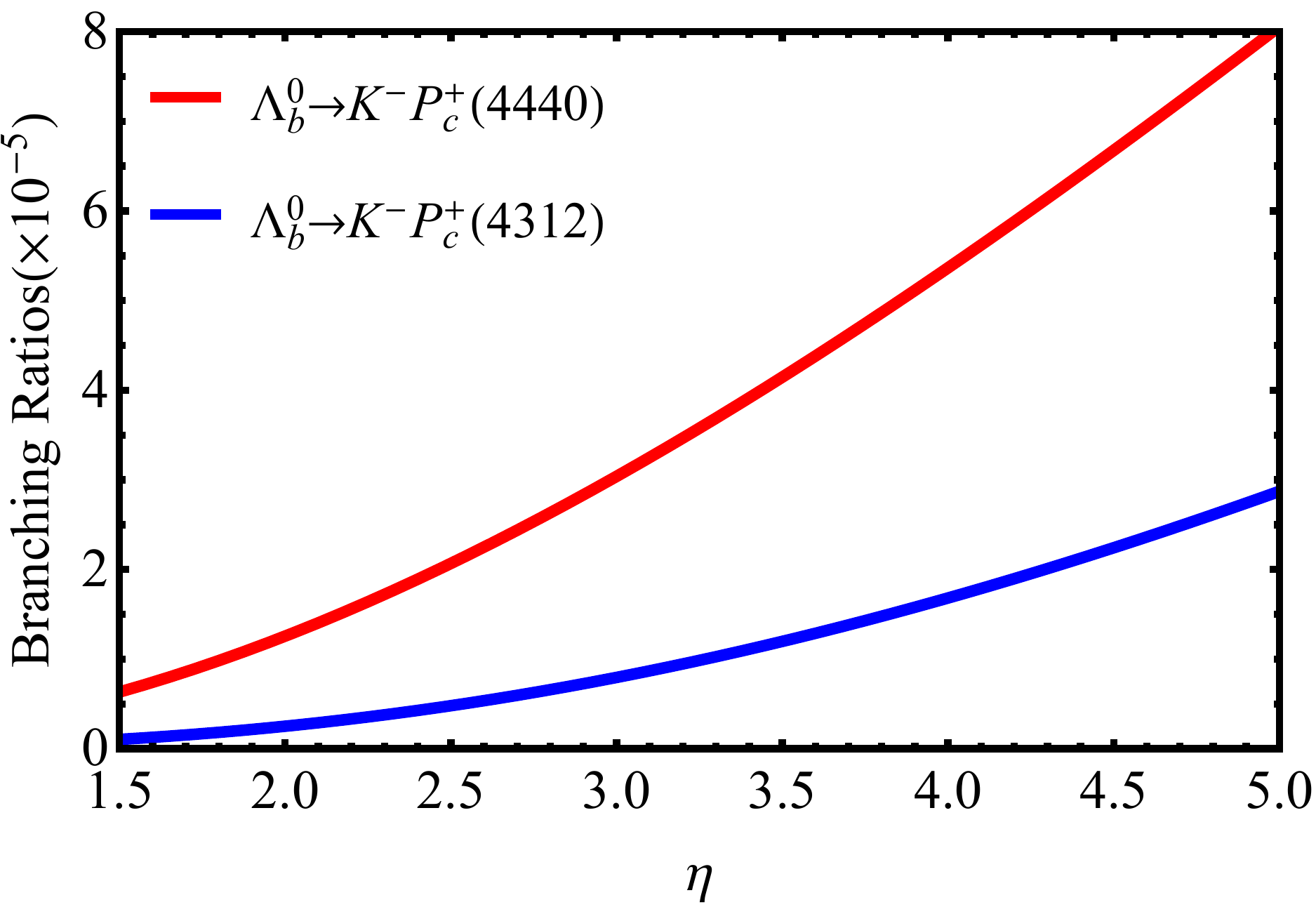}
		\caption{Similar to Fig.\ref{fig:3} but for $B^-\to \overline{\Lambda}_c^-\Xi_c^{(\prime)0}$ and $\Lambda_b^0\to P_c^+K^-$.}
		\label{fig:6}
	\end{figure}

	Considering the strong coupling constants in Refs.\cite{Nieves:2019jhp} and \cite{Lin:2019qiv}, the branching fractions of $B^-\to \overline{\Lambda}_c^-\Xi_c^{(\prime)0}$ and $\Lambda_b^0\to K^-P_c^+$ are obtained by the rescattering mechanism with the results shown in Fig.\ref{fig:6}. It can be seen that the experimental results of $Br(B^-\to \overline{\Lambda}_c^-\Xi_c^{0})=(0.95\pm 0.23)\times10^{-3}$ \cite{PDG} and $Br(B^-\to \overline{\Lambda}_c^-\Xi_c^{\prime0})=(0.34\pm 0.20)\times10^{-3}$ \cite{Li:2019glu} can be revealed at around $\eta\approx3.0$. For $\Lambda_b^0\to K^-P_c^+$, we only consider the $J^P=1/2^-$ states $P_c^+(4312)$ and $P_c^+(4440)$ for convenience. It can be seen from Fig.\ref{fig:6} that the branching fractions of $\Lambda_b^0\to K^-P_c^+$ are at the order of $10^{-5}$, similarly to those of $B^-\to D^-X_{0,1}$. The difference between $\Lambda_b^0\to K^-P_c^+$ and $B^-\to D^-X_{0,1}$ is the spectators which is an antiquark in $B^-$ decays whereas a  diquark in $\Lambda_b^0$ decays, compared with Fig.\ref{fig:1DXtopo} and Fig.\ref{fig:5}. The analogy can also be seen in $Br(\Lambda_b^0\to \Lambda_c^+D_s^-)=(1.10\pm0.10)\%\approx Br(B^-\to D^0D_s^-)=(0.90\pm0.09)\%$ \cite{PDG}. Therefore, it can be expected that $\Lambda_b^0\to K^-P_c^+$ and $B^-\to D^-X_{0,1}$ have similar branching fractions. 
	
	\section{Predictions on $B^-\to\pi^-X_{0,1}$}\label{sec:BtopiX}
\begin{figure}[bp]
		\centering
		\subfigure[]{
			\begin{minipage}[]{0.32\linewidth}
				\includegraphics[scale=0.22]{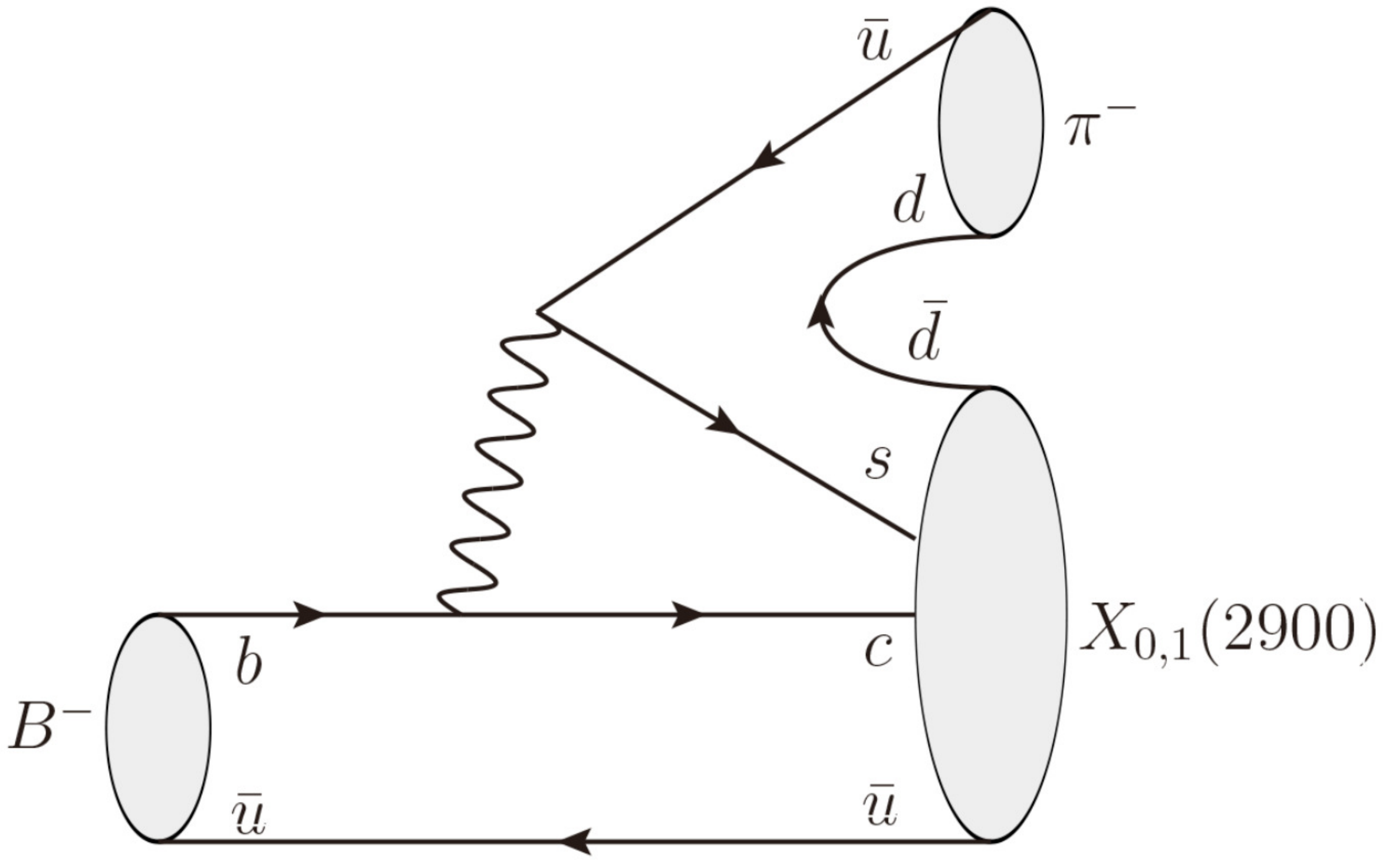}
				\label{fig:7a}
			\end{minipage}
		}
		\subfigure[]{
			\begin{minipage}[]{0.35\linewidth}
				\includegraphics[scale=0.22]{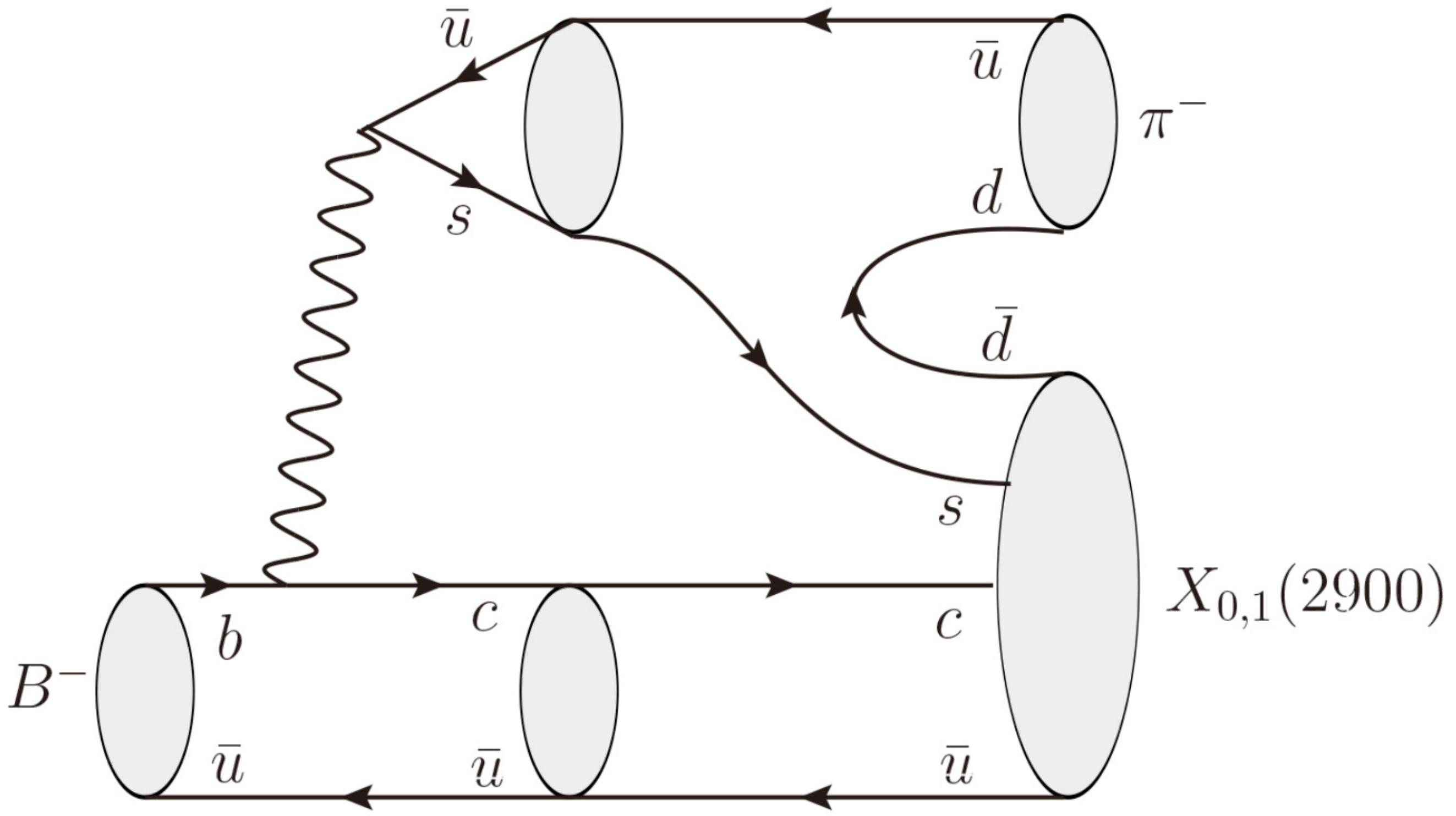}
				\label{fig:7b}
			\end{minipage}
		}
		\subfigure[]{
			\begin{minipage}[]{0.28\linewidth}
				\includegraphics[scale=0.18]{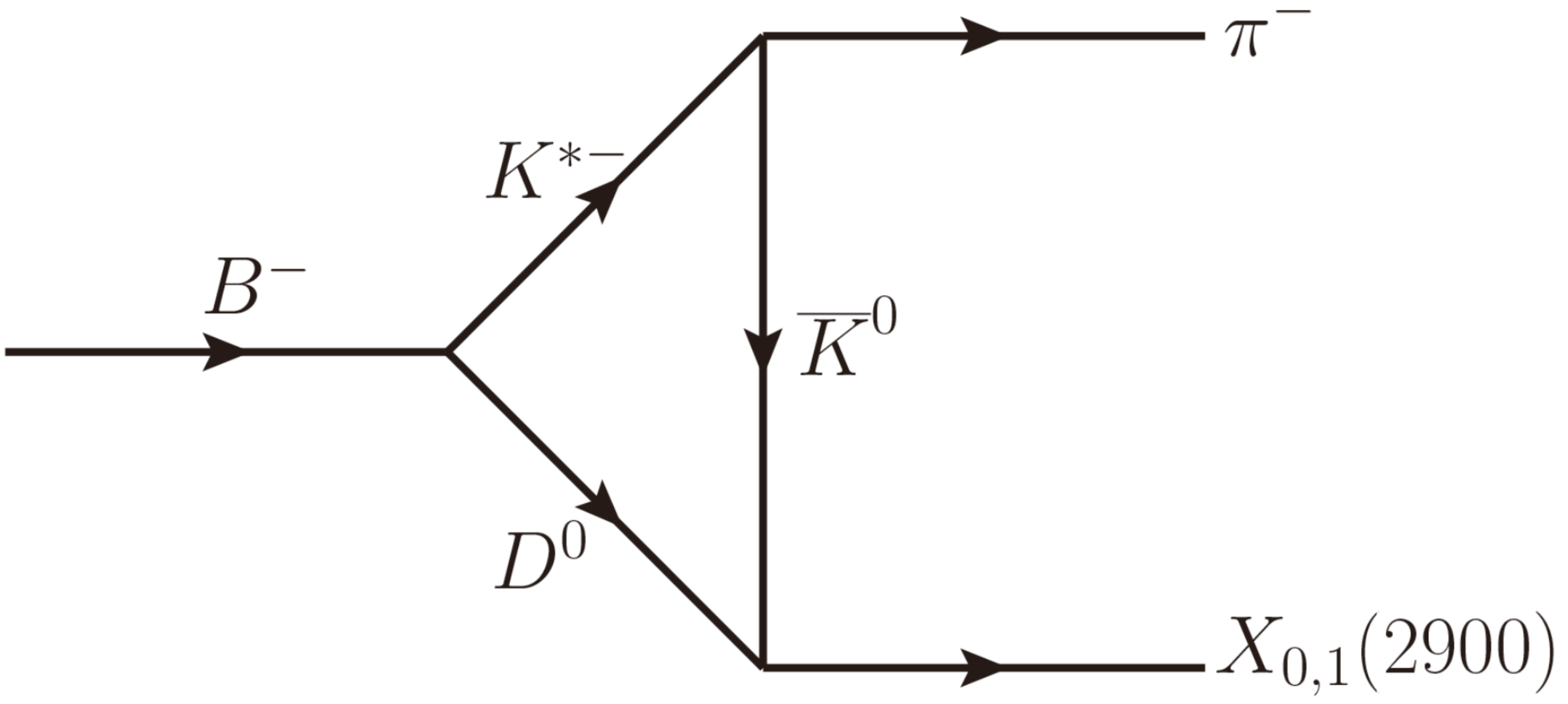}
				\label{fig:7c}
			\end{minipage}
		}
		\caption{Similar to Fig.\ref{fig:4} but for $B^-\to \pi^-X_{0,1}(2900)$. \textcolor{diff}{Those three diagrams are very similar to $B^-\to D^-X_{0,1}$ decays with a charm anti-quark replaced by an up anti-quark.}}
		\label{fig:7}
	\end{figure}

\begin{figure}[tbp]
		\centering
		\subfigure[]{
			\begin{minipage}[]{0.32\linewidth}
				\includegraphics[scale=0.27]{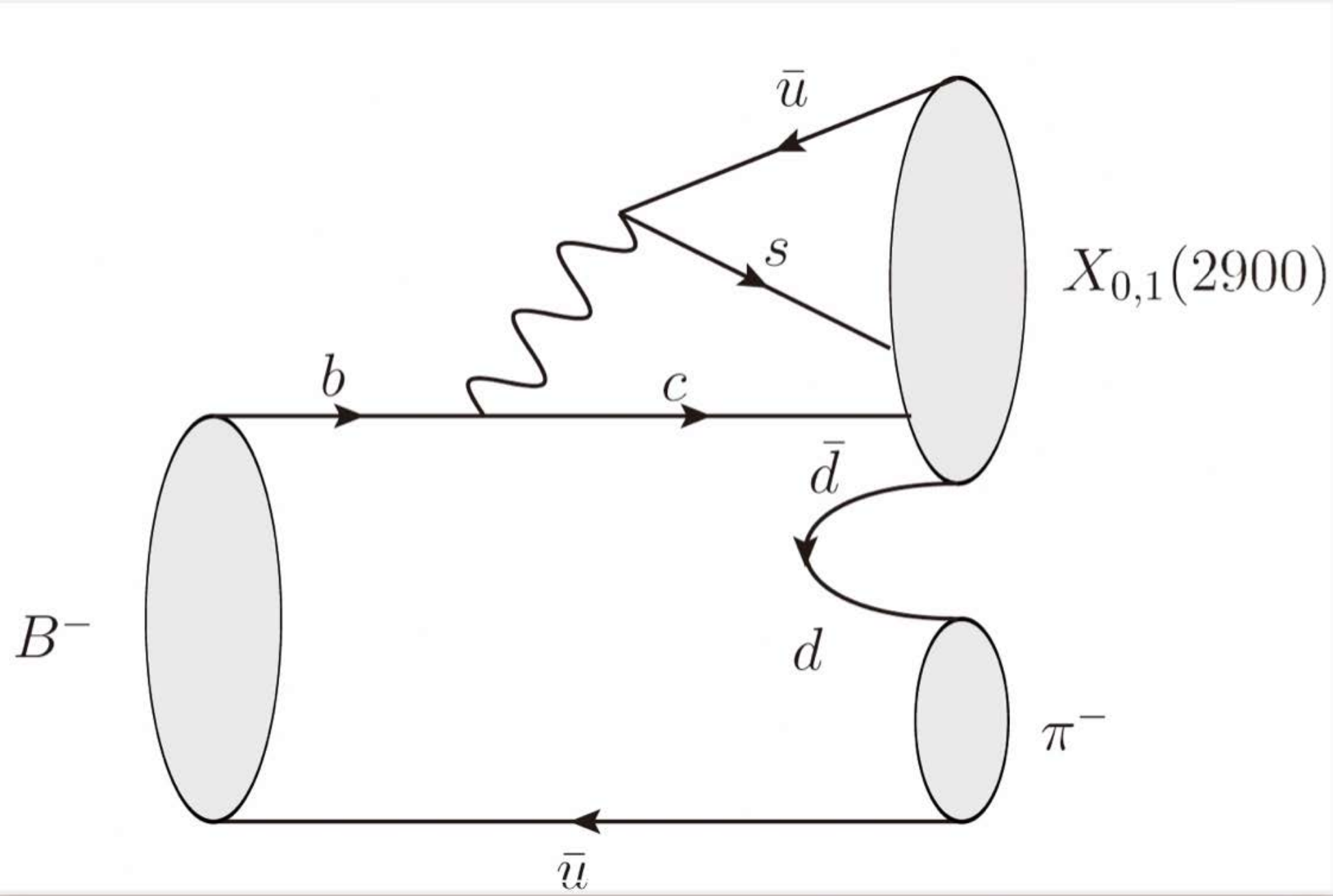}
				\label{fig:8a}
			\end{minipage}
		}
		\subfigure[]{
			\begin{minipage}[]{0.35\linewidth}
				\includegraphics[scale=0.27]{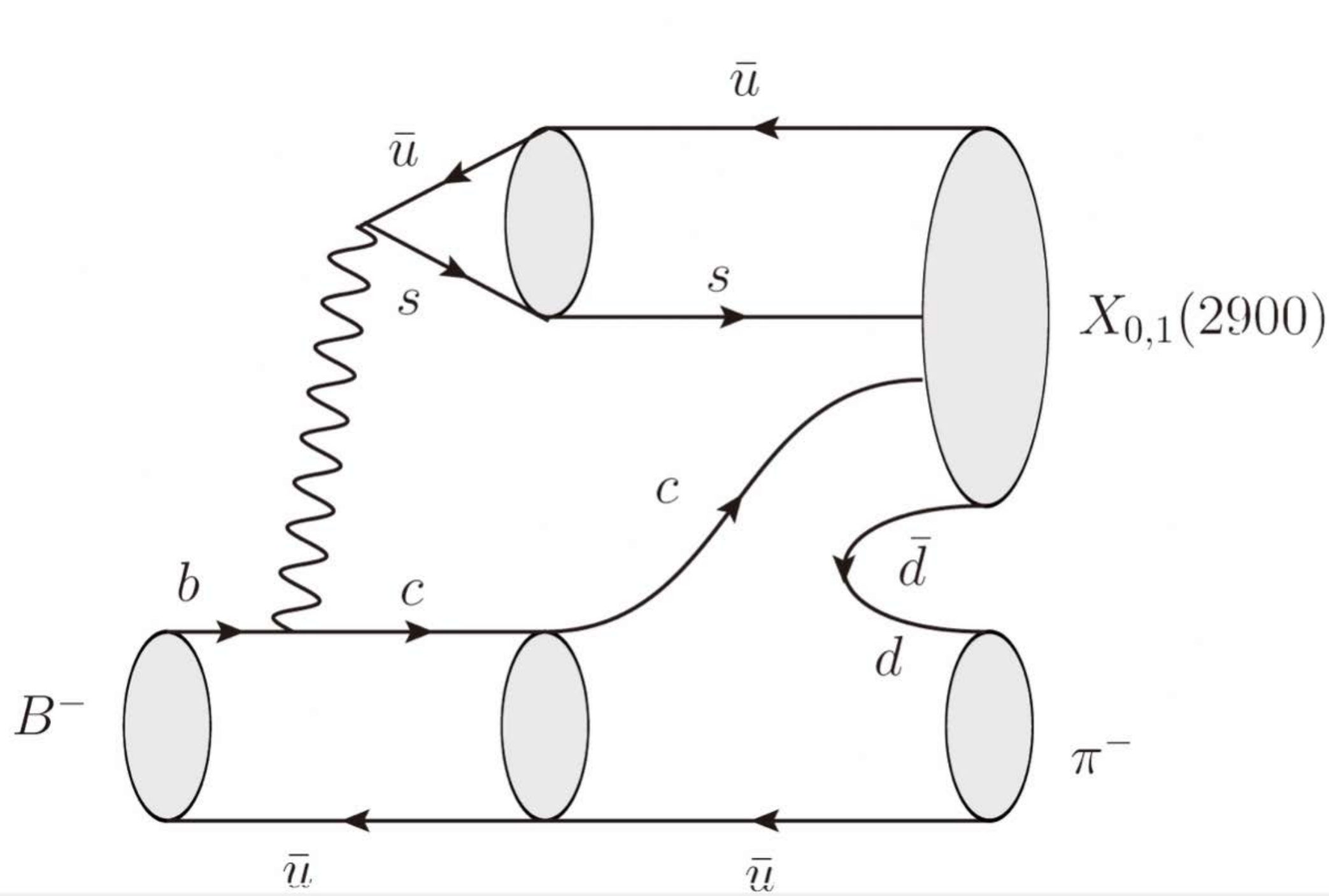}
				\label{fig:8b}
			\end{minipage}
		}
		\subfigure[]{
			\begin{minipage}[]{0.28\linewidth}
				\includegraphics[scale=0.25]{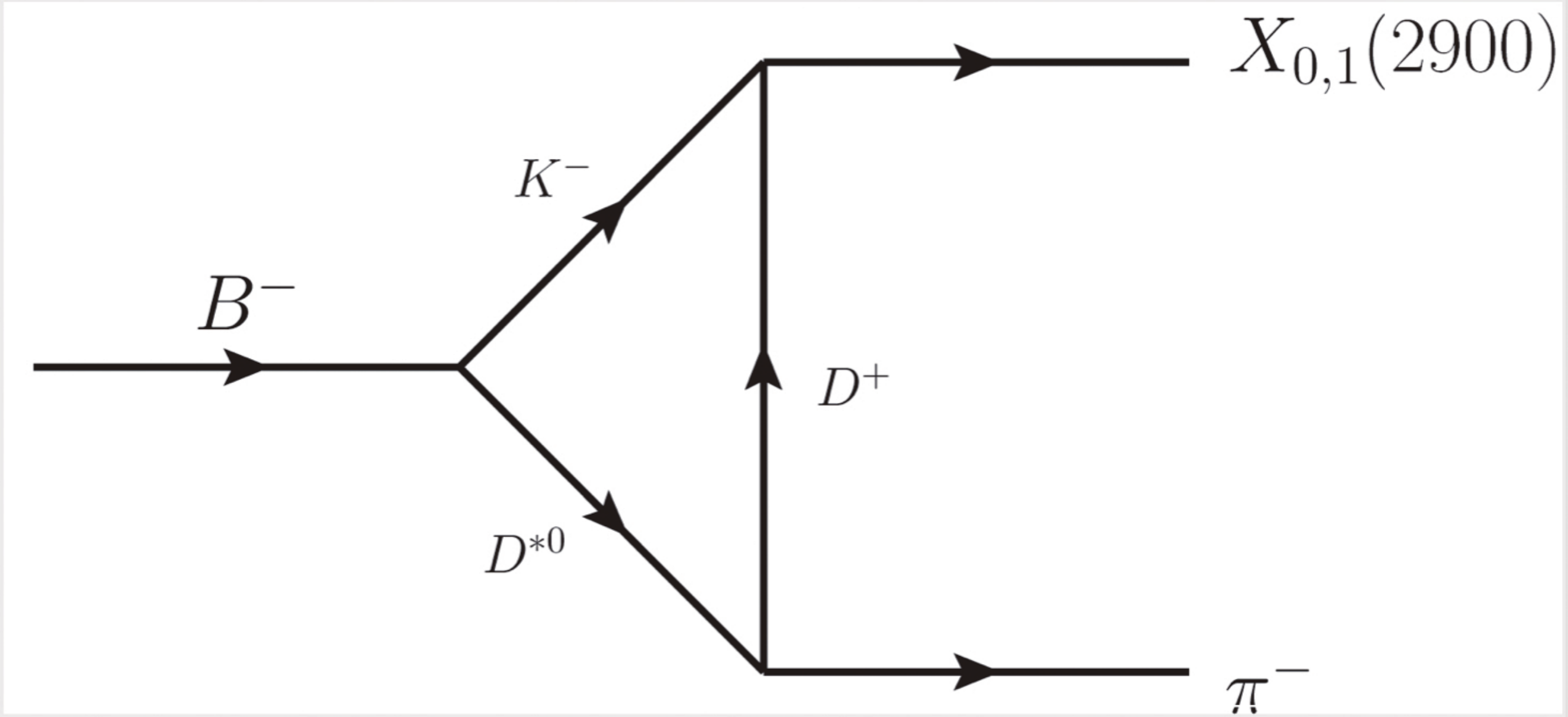}
				\label{fig:8c}
			\end{minipage}
		}
		\caption{\textcolor{diff}{The same to Fig.\ref{fig:7} but for the spectator $\overline{u}$ quark in B meson absorbed by $\pi^-$.}}
		\label{fig:8}
	\end{figure}

	Exotic states $X_{0,1}$ have alreadly been observed in $B^-\to D^-X_{0,1}$ decays, but have to be confirmed by other processes in the future experiments, for example, $B^-\to \pi^-X_{0,1}$ decays.
	\textcolor{diff}{The decays $B^-\to \pi^-X_{0,1}$ receive two different contributions. One is very similar to the $B^-\to D^-X_{0,1}$ decays with a charm anti-quark replaced by an up anti-quark. The topological, quark-level and hadron-level diagrams of $B^-\to \pi^-X_{0,1}$ are shown in Fig.\ref{fig:7}. The other contribution comes from the spectator $\overline{u}$ quark in B meson absorbed by $\pi^-$ as shown in Fig.\ref{fig:8}.}

	Theroetically speaking, compared to $B^-\to D^-X_{0,1}$ decays, the amplitudes of $B^-\to \pi^-X_{0,1}$ decays are smaller by a CKM factor of $|V_{us}|/|V_{cs}|\approx0.225$ \textcolor{diff}{The naive expectations of branching fractions are $Br(B^-\to \pi^-X_0)\sim \mathcal{O}(10^{-7})$ and $Br(B^-\to \pi^-X_1)\sim \mathcal{O}(10^{-6})$}. On the contrary, $D^-$ meson has to be reconstructed by $D^-\to K^+\pi^-\pi^-$ process in the experiments which suffers a factor of the corresponding branching fraction and another factor of around one order smaller due to two more tracks in the final states at LHCb.
	Therefore, naively speaking, $B^-\to \pi^-X_{0,1}$ processes are good alternative processes to check the results in $B^-\to D^-X_{0,1}$ decays.
	
\begin{figure}[htbp]
		\centering
		\subfigure[]{
			\begin{minipage}[]{0.48\linewidth}
				\includegraphics[width=3in]{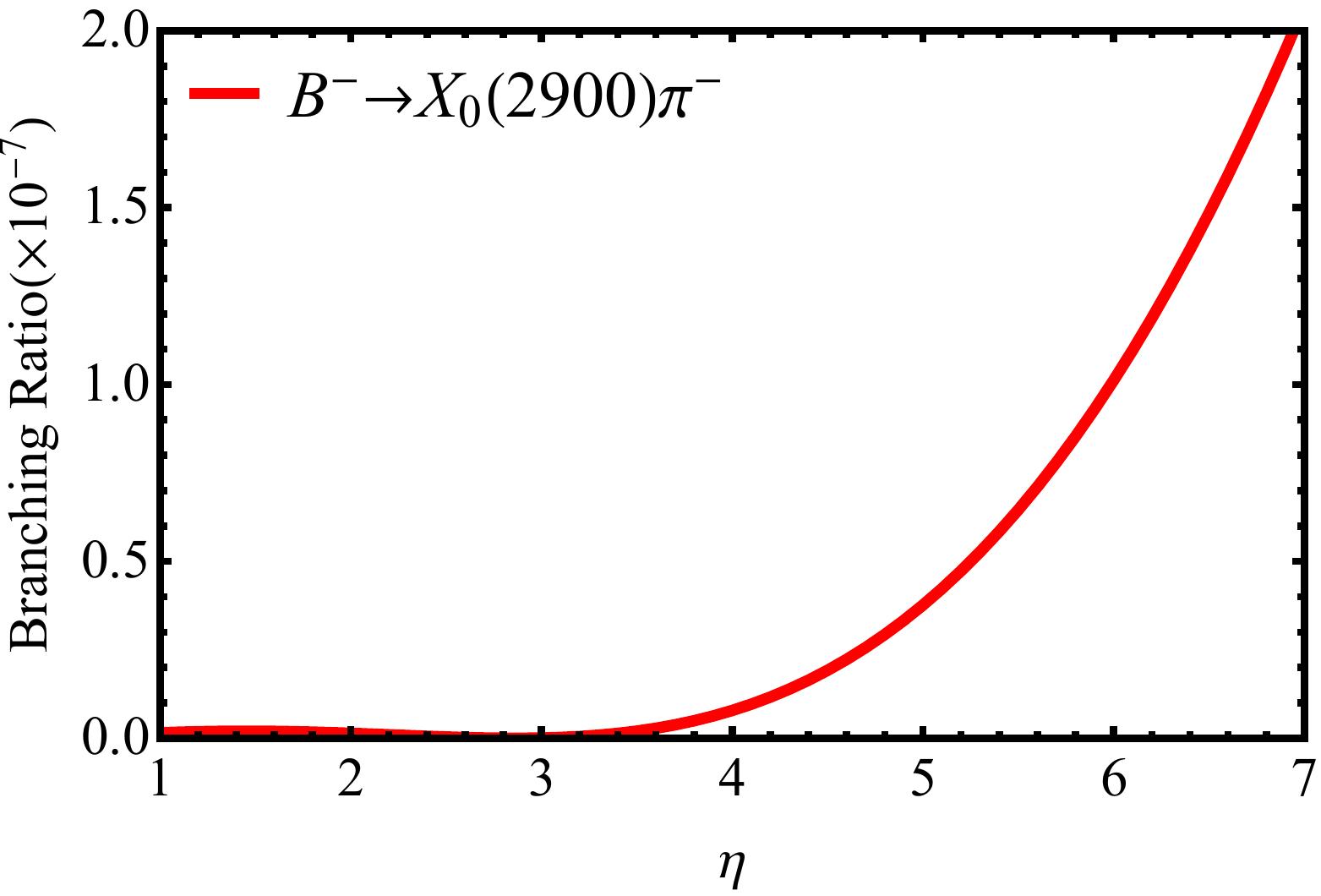}
				\label{fig:9a}
			\end{minipage}
		}
		\subfigure[]{
			\begin{minipage}[]{0.48\linewidth}
				\includegraphics[width=3in]{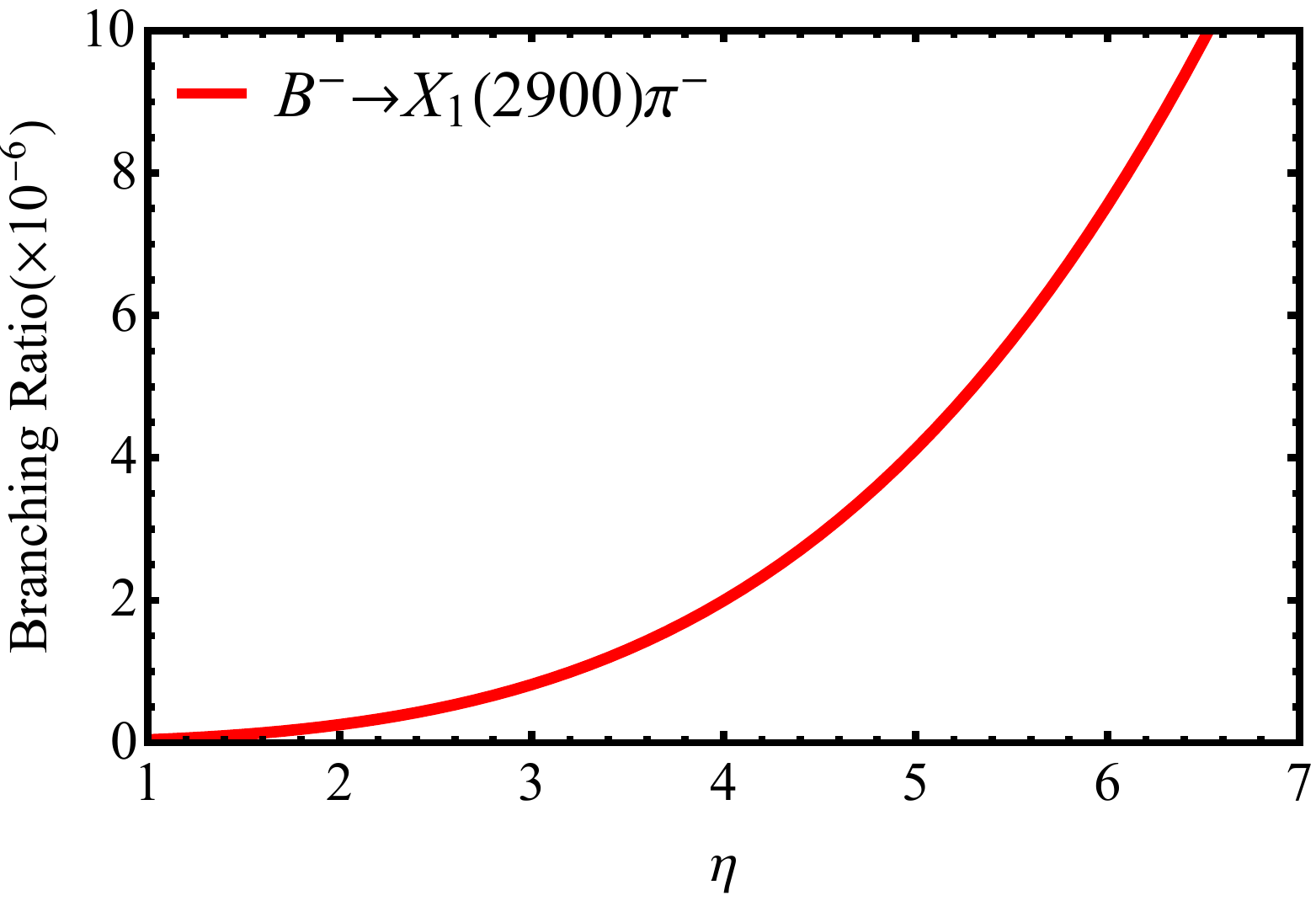}
				\label{fig:9b}
			\end{minipage}
		}
		\caption{\textcolor{diff}{(a): The theoretical branching fraction of $B^-\to\pi^- X_0(2900)$ with $\eta$ varying from 1.0 to 7.0. (b): The same as (a) but for $B^-\to\pi^- X_1(2900)$ decay.}}
		\label{fig:9}
	\end{figure}

	The numerical results of branching fraction of $B^-\to \pi^-X_{0,1}$ calculated under rescattering mechanism are shown in Fig.\ref{fig:9}. The branching fraction of $B^-\to \pi^-X_{0}$ is typically at the order of \textcolor{diff}{$10^{-8}\sim 10^{-7}$}, while the branching fraction of $B^-\to \pi^-X_{1}$ is around \textcolor{diff}{$10^{-6}$}. \textcolor{diff}{The results are comparable to the naive expectations. The contribution from Fig.\ref{fig:7}, $B^-\to K^{\ast -}D^0\to \pi^-X_{0,1}$, is much smaller than the naive expectation from $B^-\to D_s^{\ast -}D^0\to D^-X_{0,1}$ and the CKM factor. It stems from the smaller coupling constant of $g_{K^\ast K\pi}=4.6$ for the former process, and the larger $g_{D_s^\ast DK}=18.4$ for the later one, shown in the Appendix. Considering both the contributions from Fig.\ref{fig:7} and Fig.\ref{fig:8}, the branching fractions of $B^-\to \pi^-X_{0,1}$ become comparable to the naive expectations, since the coupling constants of $g_{D^\ast D\pi}=17.9$ used in Fig.\ref{fig:8} is similar to $g_{D_s^\ast DK}$.} 

	Considering $Br(B^-\to D^+K^-\pi^-)=(7.7\pm0.5)\times10^{-5}$, \textcolor{diff}{the fraction of $Br(B^-\to \pi^-X_1)/BR(B^-\to D^+K^-\pi^-)$ is $\mathcal{O}(1\%\sim 10\%)$ which is large enough to be observed.} 
	Actually, in the amplitude analysis of $B^-\to D^+K^-\pi^-$ by the LHCb collaboration \cite{Aaij:2015vea}, there is no significant enhancement or peak in the $D^+K^-$ mass spectrum, \textcolor{diff}{due to the limited data of $3fb^{-1}$. $B^-\to \pi^-X_1(2900)$ is thus expected to be observed by more data.}
	
	\section{Isospin analysis on $B\to DX_{0,1}$}\label{sec:Isospin}
	The quark flovors of $X_{0,1}(2900)$ are $cs\bar{u}\bar{d}$, but their isoapin are not determined. Ref.\cite{Karliner:2020vsi,Hu:2020mxp,Liu:2020nil,He:2020btl,Xue:2020vtq} predict that the $X_0(2900)$ is a isospin singlet, while Ref.\cite{He:2020btl} also considers $X_1(2900)$ as isospin singlet.
	
	A state of $cs\bar{u}\bar{d}$ could either be isospin singlet or triplet. In order to investigate their isospins, we could study some other processes of $B\to DX_{0,1}$. The isospins of $X_0$ and $X_1$ are not necessarily the same. Therefore the following discussions can be used for each of $X_0$ and $X_1$, labeled as $X_i$.
	
	In the decay of $B^-\to D^-X_i$, the weak interaction happens as $b\to c\bar{c}s$, which does not change the isospin. As the initial states of $B$ mesons are isospin doublet, the final states of $DX_i$ must be isospin doublet as well. In the following, we will discuss the cases of $X_i$ as isospin-0 or isospin-1 states, respectively.
	
	Firstly, in the case that  $X_i$ is an isospin singlet state, the isospin of $DX_i$ must be $1/2$ and the same as the initial state of $B$ mesons. Then
	\begin{equation}
		\mathcal{A}(\overline{B}^0\to \overline{D}^0X_i^0)=\mathcal{A}(B^-\to D^-X_i^0).
	\end{equation}
	The branching fractions of the above processes are equal to each other. With $Br(\overline{B}^0\to \overline{D}^0D^+K^-)=(1.07\pm 0.11)\times10^{-3}$, we then have
	\begin{equation}
		Br(\overline{B}^0\to \overline{D}^0X_0^0)/Br(\overline{B}^0\to \overline{D}^0D^+K^-)=(1.15\pm0.38)\%,
	\end{equation}
or
	\begin{equation}
		Br(\overline{B}^0\to \overline{D}^0X_1^0)/Br(\overline{B}^0\to \overline{D}^0D^+K^-)=(6.29\pm2.11)\%.
	\end{equation}
These ratios would be helpful for the experimental analysis. 
	
	In the second case that the isospin of $X_i^0$ is one, the isospin of the final state $DX_i$ is a linear combination of isospin-1/2 and isospin-3/2. For example,
	\begin{equation}
		\begin{aligned}
		|D^-X_i^0\rangle=&\sqrt{\frac{2}{3}}|\frac{3}{2},-\frac{1}{2}\rangle+\frac{1}{\sqrt{3}}|\frac{1}{2},-\frac{1}{2}\rangle\\
		|\overline{D}^0X_i^-\rangle=&\frac{1}{\sqrt{2}}|\frac{3}{2},-\frac{1}{2}\rangle-\sqrt{\frac{2}{3}}|\frac{1}{2},-\frac{1}{2}\rangle\\
		|\overline{D}^0X_i^0\rangle=&\sqrt{\frac{3}{2}}|\frac{3}{2},\frac{1}{2}\rangle-\sqrt{\frac{1}{3}}|\frac{1}{2},\frac{1}{2}\rangle\\
		|D^-X_i^+\rangle=&\frac{1}{\sqrt{3}}|\frac{3}{2},\frac{1}{2}\rangle+\sqrt{\frac{2}{3}}|\frac{1}{2},\frac{1}{2}\rangle
		\end{aligned}
	\end{equation}
	Then we have
	\begin{equation}
		\begin{aligned}
		\mathcal{A}(B^-\to D^-X_i^0)=&\frac{1}{\sqrt{3}}A_{\frac{1}{2}}\\
		\mathcal{A}(B^-\to \overline{D}^0X_i^-)=&-\sqrt{\frac{2}{3}}A_{\frac{1}{2}}\\
		\mathcal{A}(\overline{B}^0\to \overline{D}^0X_i^0)=&-\sqrt{\frac{1}{3}}A_{\frac{1}{2}}\\
		\mathcal{A}(\overline{B}^0\to D^-X_i^+)=&\sqrt{\frac{2}{3}}A_{\frac{1}{2}}\\
		\end{aligned}
	\end{equation}
where $A_{\frac{1}{2}}$ is the isospin amplitude in these decays.
	Therefore the isospin relations of the amplitudes are
	\begin{equation}
		\begin{aligned}
		\mathcal{A}(B^-\to\overline{D}^0X_i^-)=&-\sqrt{2}\mathcal{A}(B^-\to D^-X_i^0)
		=-\mathcal{A}(\overline{B}^0\to D^-X_i^+)=\sqrt{2}\mathcal{A}(\overline{B}^0\to\overline{D}^0X_i^0)
		\end{aligned}
	\end{equation}

	\begin{figure}[htbp]
		\centering
		\subfigure[]{
			\begin{minipage}[]{0.45\linewidth}
				\includegraphics[width=2.0in]{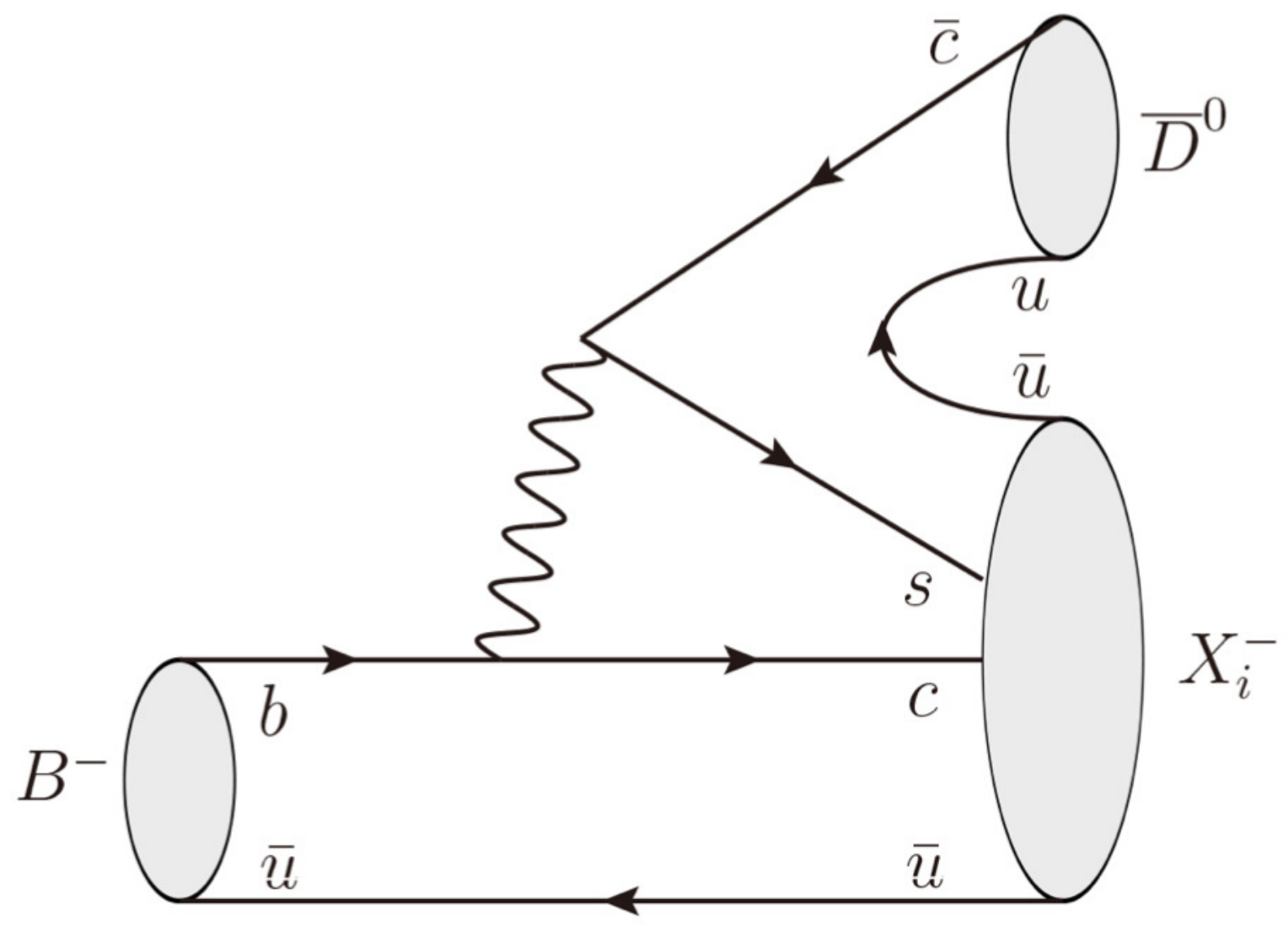}
				\label{fig:11a}
			\end{minipage}
		}
		\subfigure[]{
			\begin{minipage}[]{0.45\linewidth}
				\includegraphics[width=2.0in]{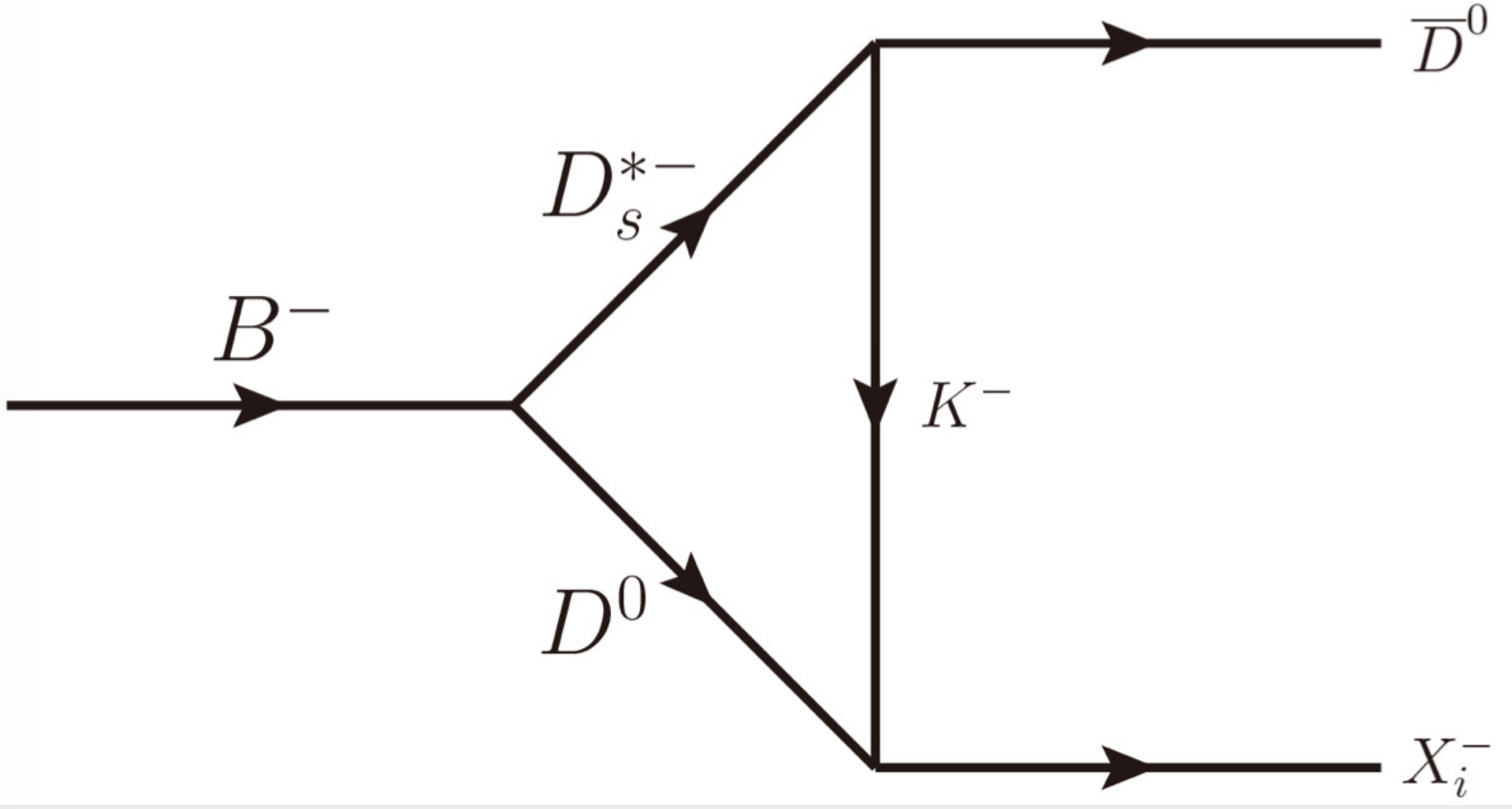}
				\label{fig:11b}
			\end{minipage}
		}
		\subfigure[]{
			\begin{minipage}[]{0.45\linewidth}
				\includegraphics[width=2.0in]{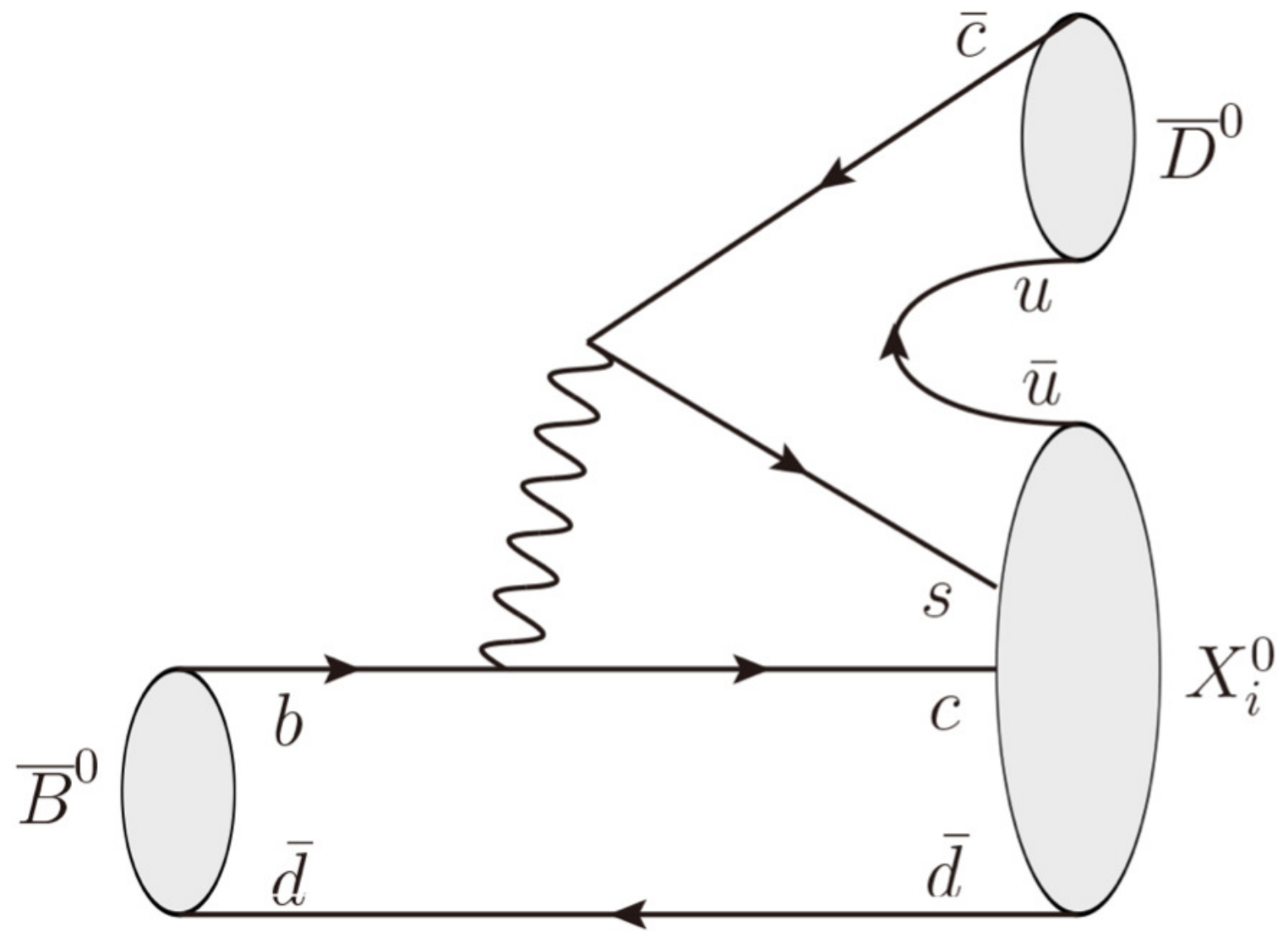}
				\label{fig:11c}
			\end{minipage}
		}
		\subfigure[]{
			\begin{minipage}[]{0.45\linewidth}
				\includegraphics[width=2.0in]{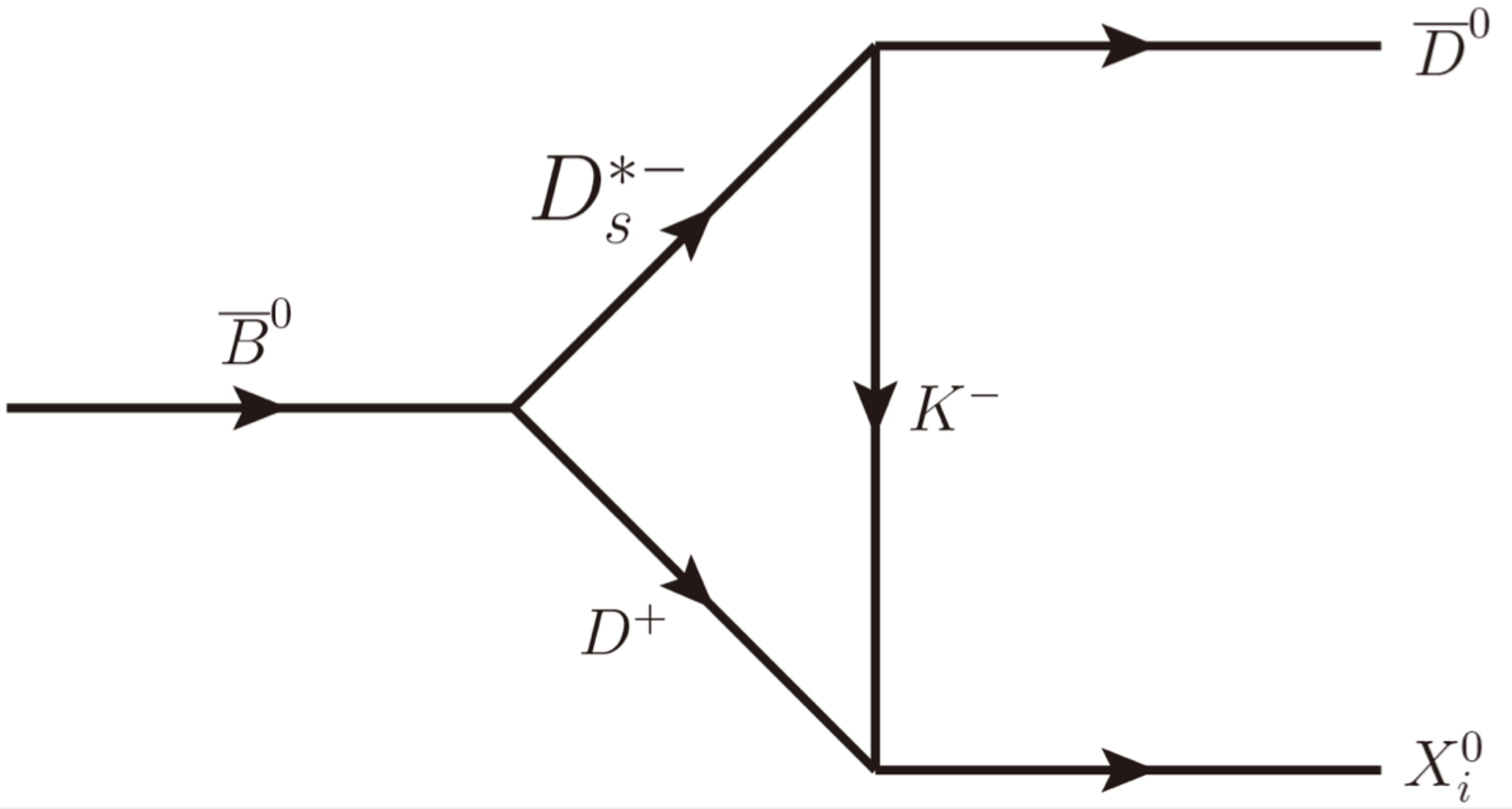}
				\label{fig:11d}
			\end{minipage}
		}
		\subfigure[]{
			\begin{minipage}[]{0.45\linewidth}
				\includegraphics[width=2.0in]{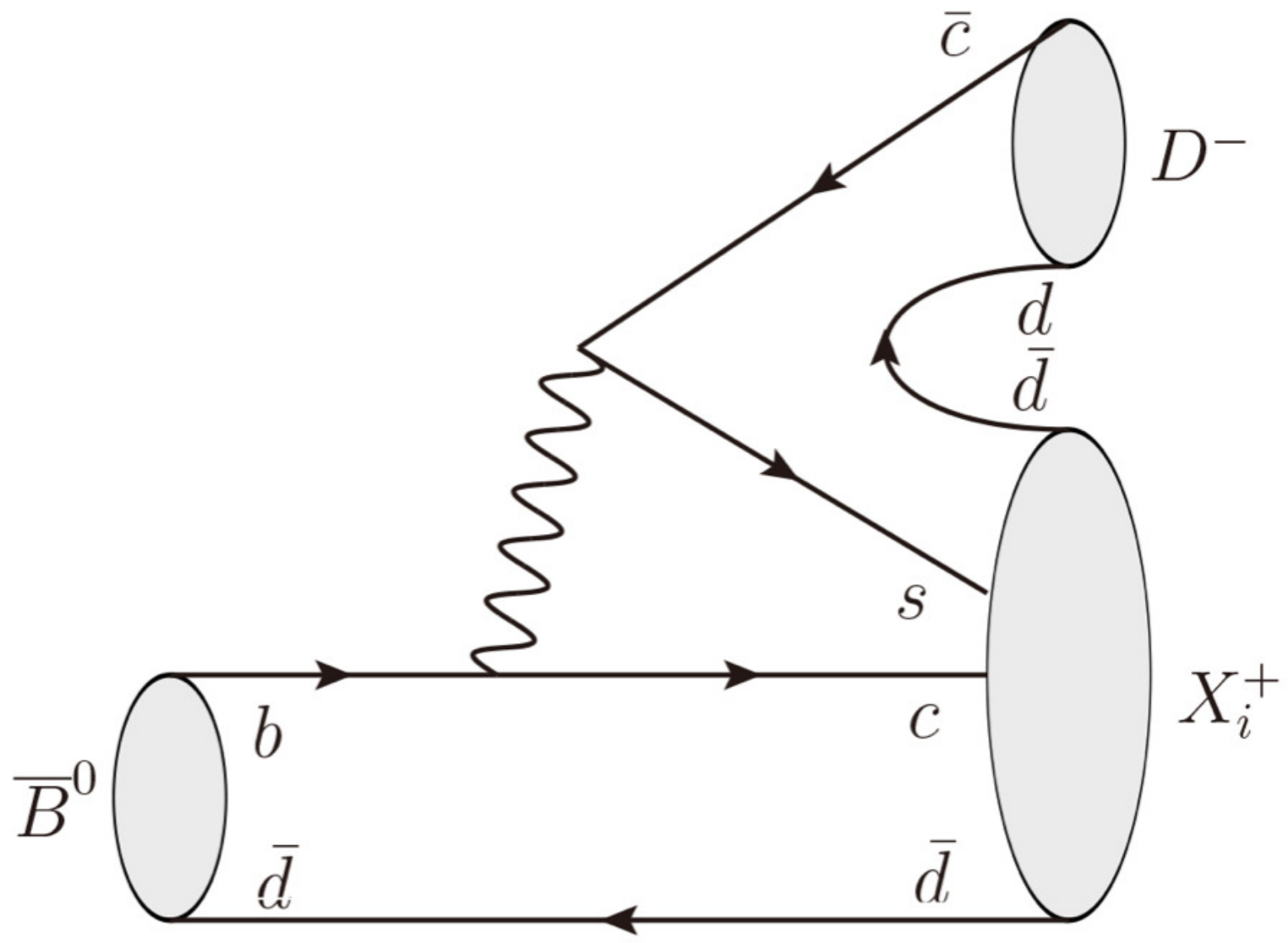}
				\label{fig:11e}
			\end{minipage}
		}
		\subfigure[]{
			\begin{minipage}[]{0.45\linewidth}
				\includegraphics[width=2.0in]{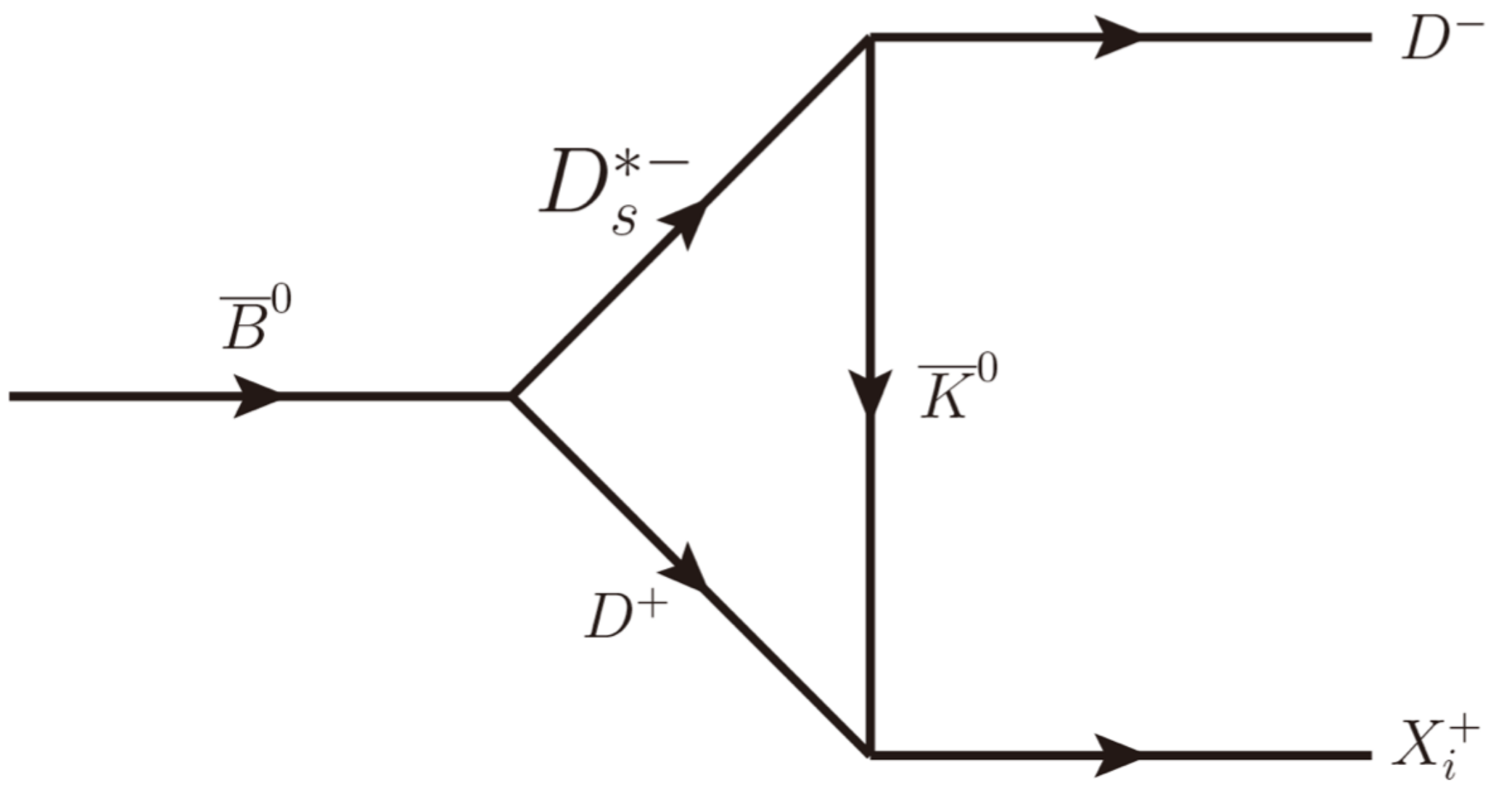}
				\label{fig:11f}
			\end{minipage}
		}

		\caption{The topological and hadronic triangle diagrams of $B^-\to \overline{D}^0X_i^-$, $\overline{B}^0\to \overline{D}^0X_i^0$ and $\overline{B}^0\to D^-X_i^+$, respectively.}
		\label{fig:11}
	\end{figure}

	The factor of $\sqrt{2}$ in the above relation can also be understood by the rescattering mechanism. The strong couplings of $X_iDK$ have the following relations under the SU(3) flavor symmetry \cite{He:2020jna}
	\begin{equation}
		g_{X_i^-D^0K^-}=g_{X_i^+D^+\overline{K}^0}=\sqrt{2}g_{X_i^0D^0\overline{K}^0}=\sqrt{2}g_{X_i^0D^+K^-}
	\end{equation}
	With $Br(B^-\to D^0\overline{D}^0K^-)= (1.45\pm0.33)\times10^{-3}$ and $Br(\overline{B}^0\to D^+D^-\overline{K}^0)=(7.5\pm 1.7)\times 10^{-4}$ \cite{PDG}, we have
	\begin{equation}
		\begin{aligned}
		Br(B^-\to\overline{D}^0X_0^-)/Br(B^-\to D^0\overline{D}^0K^-)=&(1.8\pm0.6)\%\\
		Br(\overline{B}^0\to D^-X_0^+)/Br(\overline{B}^0\to D^+D^-\overline{K}^0)=&(3.2\pm1.2)\%
		\end{aligned}
	\end{equation}
or 
	\begin{equation}
		\begin{aligned}
		Br(B^-\to\overline{D}^0X_1^-)/Br(B^-\to D^0\overline{D}^0K^-)=&(9.2\pm3.6)\%\\
		Br(\overline{B}^0\to D^-X_1^+)/Br(\overline{B}^0\to D^+D^-\overline{K}^0)=&(18.0\pm7.0)\%
		\end{aligned}
	\end{equation}
	which are useful for the experimental measurements.
	
	$\overline{B}^0\to\overline{D}^0X_i^0$ has the same branching fraction in both cases of isospin-0 and 1 states of $X_i^0$. It can not be used to test the isospin of $X_i^0$, but to confirm the existence of $X_i^0$. If $B^-\to \overline{D}^0X_i^-$ or $\overline{B}^0\to D^-X_0^+$ were observed, the isospin would be determined to be one.
	
	\section{Summary}\label{sec:Summary}
	In this article, motivated by the observation of the exotic states $X_{0,1}(2900)$, we calculate the branching fractions of $B^-\to D^-X_{0,1}$ using rescattering mechanism, which are consistant with experimental measurements. The rescattering mechanism is tested by the processes $B^-\to \overline{\Lambda}_c^-\Xi_c^{(\prime)0}$ and $\Lambda_b^0\to P_c^+K^-$. The branching fractions of $B^-\to \pi^-X_{0,1}$ are predicted with large uncertainties. Finally, the isospins of $X_{0,1}(2900)$ are discussed. If $B^-\to \overline{D}^0X_i^-$ or $\overline{B}^0\to D^-X_0^+$ were observed, $X_i(2900)$ could be determined as an isospin triplet state.
	
	\section*{Acknowlegdements}
	This work is partly supported by the National Nature Science Foundation of China under the Grant No. 11775117, U1732101, 11975112 and 11705056, by Gansu Natural Science Fund under grant No.18JR3RA265, and by the Fundamental Research Funds for the Central Universities under Grant No. lzujbky-2019-55.
	
	\appendix
	
	\section*{Appendix: Lagrangian and couplings}
	The effective Lagrangians used in the rescattering mechanism are: \cite{Yu:2017zst,Lin:2019qiv,Ahmed:2001xc}
	\begin{equation}
		\begin{aligned}
		\mathcal{L}_{D^\ast DP}=&-i g_{D^\ast DP}(D^i \partial^\mu P_{ij} D_\mu^{\ast j \dagger} - D_\mu^{\ast i}\partial^\mu P_{ij} D^{j\dagger})\\
		\mathcal{L}_{\overline{D}\Sigma_cP_c}=& g_{\overline{D}\Sigma_cP_c}\Sigma_c\overline{P}_c\overline{D}\\
		\mathcal{L}_{\Lambda_c ND_q}=& g_{\Lambda_c ND_q}(\overline{\Lambda}_c i\gamma_5 D_q N + h.c.)\\
		\mathcal{L}_{\Lambda_c ND_q^\ast}=& f_{1\Lambda_c ND_q^\ast}(\overline{\Lambda}_c \gamma_\mu D_q^\ast N+h.c.)+\frac{f_{2\Lambda_c ND_q^\ast}}{m_{\Lambda_c}+m_N}(\overline{\Lambda}_c\sigma_{\mu\nu}\partial^\mu D_q^{\ast \nu} N+h.c.)\\
		\mathcal{L}_{P \mathcal{B}\mathcal{B}}=&g_{P \mathcal{B}\mathcal{B}}Tr\big[\overline{\mathcal{B}} i\gamma_5 P \mathcal{B}\big]\\
		\mathcal{L}_{V \mathcal{B}\mathcal{B}}=&f_{1V \mathcal{B}\mathcal{B}}Tr\big[\overline{\mathcal{B}} \gamma_\mu V^\mu \mathcal{B}\big] + \frac{f_{2V \mathcal{B}\mathcal{B}}}{2m_\mathcal{B}}Tr\big[\overline{\mathcal{B}}\sigma_{\mu\nu}\partial^\mu V^{\ast \nu}\mathcal{B}\big]\\
		\mathcal{L}_{SPP}=&-g_{SPP}m_s SPP\\
		\mathcal{L}_{VPP}=&ig_{VPP}Tr\big[V^\mu[P,\partial_\mu P]\big]
		\end{aligned}
	\end{equation}
	where the corresponding $P$, $V$ and $\mathcal{B}$ represent the matrices, respectively
	\begin{equation}
	\begin{aligned}
		P=\left(
		\begin{array}{ccc}
		\frac{\pi^0}{\sqrt{2}}+\frac{\eta_8}{\sqrt{6}}&\pi^+&K^+\\
		\pi^-&-\frac{\pi^0}{\sqrt{2}}+\frac{\eta_8}{\sqrt{6}}&K^0\\
		K^-&\bar{K}^0&-\sqrt{\frac{2}{3}}\eta_8\\
		\end{array}\right)&+\frac{1}{\sqrt{3}}\left(
		\begin{array}{ccc}
		\eta_1&0&0\\
		0&\eta_1&0\\
		0&0&\eta_1\\
		\end{array}\right)\\
		V=\left(		\begin{array}{ccc}
		\frac{\rho^0}{\sqrt{2}}+\frac{\omega}{\sqrt{2}}&\rho^+&K^{\ast +}\\
		\rho^-&-\frac{\rho^0}{\sqrt{2}}+\frac{\omega}{\sqrt{2}}&K^{\ast 0}\\ 		K^{\ast -}&\bar{K}^{\ast 0}&\phi\\
		\end{array}\right)\ \ 
		&\mathcal{B}=\left(
		\begin{array}{ccc}
		\frac{\Sigma^0}{\sqrt{2}}+\frac{\Lambda}{\sqrt{6}}&\Sigma^+&p\\
		\Sigma^-&-\frac{\Sigma^0}{\sqrt{2}}+\frac{\Lambda}{\sqrt{6}}&n\\
		\Xi^-&\Xi^0&-\frac{2}{\sqrt{6}}\Lambda
		\end{array}
		\right)
		\end{aligned}
	\end{equation}
	
	The strong coupling constants are taken from the literature \cite{Yu:2017zst,Lin:2019qiv,Ahmed:2001xc}, and listed in Table\ref{table:app}.
	\begin{table}
		\caption{The strong coupling constants.}
		\begin{tabular}{cccccc}
			\hline
			\hline
			\ \ \ \ \ \ \ \ \ \ \ \ \ \ \ \ \ \ vertex\ \ \ \ \ \ \ \ \ \ \ \ \ \ \ \ \ \ &g&\ \ \ \ \ \ \ \ \ \ \ \ \ \ \ \ \ \ vertex\ \ \ \ \ \ \ \ \ \ \ \ \ \ \ \ \ \ &g&\ \ \ \ \ \ \ \ \ \ \ \ \ \ \ \ \ \ vertex\ \ \ \ \ \ \ \ \ \ \ \ \ \ \ \ \ \ &g\\
			\hline
			$\Lambda_c^+\to \Lambda^0D_s^+$&5.83&$\Lambda_c^+\to \Sigma^0D_s^+$&9.31&$\Xi_c^{\prime 0}\to \Lambda^0D^0$&6.43\\
			\hline
			$\Xi_c^{\prime 0}\to \Sigma^0D^0$&3.71&$X_0(2900)\to \overline{K}^0D^0$&1.0&$X_1(2900)\to \overline{K}D^0$&9.3\\
			\hline
			$D_s^{\ast -}\to \overline{K}^0D^-$&18.4&$P_c^+(4312)\to \Lambda_c^+D^0$&0.088&$P_c^+(4312)\to \Lambda_c^+D^{\ast 0}$&0.58\\
			\hline
			$P_c^+(4440)\to \Lambda_c^+D^0$&0.80&$P_c^+(4440)\to \Lambda_c^+D^{\ast 0}$&0.68&$\Xi_c^0\to \Lambda^0 D^0$&1.59\\
			\hline
			$\Xi_c^0\to \Sigma^0 D^0$&2.75&$K^{\ast -}\to \overline{K}^0\pi^-$&4.60&\textcolor{diff}{$D^{\ast 0}\to D^+\pi^-$}&\textcolor{diff}{17.9}\\
			\hline
		\end{tabular}
		\begin{tabular}{cccccc}
			\hline
			\ \ \ \ \ \ \ \ \ \ \ \ \ \ \ \ \ \ \ vertex\ \ \ \ \ \ \ \ \ \ \ \ \ \ \ \ \ \ \ &$f_1$&\ \ \ \ \ \ \ \ \ \ \ \ \ \ \ \ \ \ $f_2$&\ \ \ \ \ \ \ \ \ \ \ \ \ \ \ \ \ \ \ vertex\ \ \ \ \ \ \ \ \ \ \ \ \ \ \ \ \ \ \ &$f_1$&\ \ \ \ \ \ \ \ \ \ \ \ \ \ \ \ \ \ $f_2$\\
			\hline
			$\Lambda_c^+\to \Lambda^0 D_s^{\ast -}$&2.05&\ \ \ \ \ \ \ \ \ \ \ \ \ \ \ \ \ \ 7.78&$\Xi_c^{\prime 0}\to \Lambda^0D^{\ast 0}$&-5.63&\ \ \ \ \ \ \ \ \ \ \ \ \ \ \ \ \ \ -10.0\\
			\hline
			$\Lambda_c^+\to \Sigma^0 D_s^{\ast -}$&3.55&\ \ \ \ \ \ \ \ \ \ \ \ \ \ \ \ \ \ 13.5&$\Xi_c^{\prime 0}\to \Sigma^0D^{\ast 0}$&-3.2&\ \ \ \ \ \ \ \ \ \ \ \ \ \ \ \ \ \ -6.0\\
			\hline
			$\Xi_c^{0}\to \Lambda^0D^{\ast 0}$&3.55&\ \ \ \ \ \ \ \ \ \ \ \ \ \ \ \ \ \ 13.5&$\Xi_c^{0}\to \Sigma^0D^{\ast 0}$&2.05&\ \ \ \ \ \ \ \ \ \ \ \ \ \ \ \ \ \ 7.78\\
			\hline
			\hline
		\end{tabular}
	\label{table:app}
	\end{table}
	

\end{document}